# Vendor-Agnostic Joint Relaxometry and Myelin Water Fraction Mapping with $B_1$ and Motion Correction

Unay Dorken Gallastegi[†,1,2], Shohei Fujita[†,1,2,3,4], Yohan Jun[1,2], Antoine Delattre-Klauser[5-7], Gian Franco Piredda[5], Tom Hilbert[5-7], Cemre Ariyurek[2,8], Eugene Milshteyn[9], Shizhuo Li[1,2], Yuting Chen[1,2], Xingwang Yong[1,2], Kwok-Shing Chan[1,2], Qiang Liu[10], Seonghwan Yee[2], Yogesh Rathi[10], Maxim Zaitsev[11], Jon-Fredrik Nielsen[12], Onur Afacan[2,8], Camilo Jaimes[1,2], Patricia Ellen Grant[2,13,14], Borjan Gagoski*[2,13], Berkin Bilgic*[1,2,15]

[1]Athinoula A. Martinos Center for Biomedical Imaging, Massachusetts General Hospital, Boston, MA, United States; [2]Department of Radiology, Harvard Medical School, Boston, MA, United States; [3]Department of Radiology, Juntendo University, Tokyo, Japan; [4]Department of Radiology, The University of Tokyo, Tokyo, Japan; [5]Swiss Innovation Hub, Siemens Healthineers International AG, Lausanne, Switzerland; [6]Department of Radiology, Lausanne University Hospital and University of Lausanne, Lausanne, Switzerland; [7]LTS5, École Polytechnique Fédérale de Lausanne (EPFL), Lausanne, Switzerland; [8]Computational Radiology Laboratory, Boston Children's Hospital, Boston, MA, United States; [9]GE HealthCare, Boston, MA, USA; [10]Brigham and Women's Hospital, Harvard Medical School, Boston, Massachusetts, USA; [11]Division of Medical Physics, Department of Radiology, Faculty of Medicine, Medical Center–University of Freiburg, Freiburg, Germany; [12]Functional MRI Laboratory, Department of Radiology, University of Michigan, Ann Arbor, MI, US; [13]Fetal-Neonatal Neuroimaging and Developmental Science Center, Boston Children's Hospital, Boston, MA, United States; [14]Department of Pediatrics, Harvard Medical School, Boston, MA, United States; [15]Harvard/MIT Health Sciences and Technology, Cambridge, MA, United States

[†]U.D.G. and S.F. contributed equally to this work and share first authorship.
* B.G. and B.B. are co–senior authors.

**Corresponding author:**
Unay Dorken Gallastegi
Athinoula A. Martinos Center for Biomedical Imaging, Massachusetts General Hospital, Boston, MA, United States
Building 75, 13th Street Charlestown, MA 02129
Phone: +1-617-866-8740
Fax: +81-3-3816-0958
E-mail: udorkengallastegi@mgh.harvard.edu

## Abstract

Obtaining consistent quantitative maps of myelin content and relaxation times across different sites and vendors is essential for advancing our understanding of brain development. Herein, we present a harmonized, vendor-agnostic magnetic resonance acquisition method designed for joint $T_1$, $T_2$, and myelin water fraction mapping, along with a method for rapid $B_1^+$ and $B_1^-$ field estimation. We used our dictionary-based fitting and multi-compartment modeling for joint mapping of $T_1$, $T_2$ and myelin water fraction. Self-navigation–based retrospective motion correction was integrated with subspace reconstruction to track and correct rigid head motion during scanning, operating without the need for external hardware. Simulations, phantom and in vivo experiments confirmed the sensitivity and accuracy of the method, particularly for short $T_2$ values corresponding to myelin, and demonstrated consistent performance across multiple scanner types. Coupled with the harmonized calibration scan, the proposed package offers a practical tool for multi-site, multi-vendor neuroimaging studies in both adult and pediatric populations.

## 1. Introduction

Emerging large-scale studies of brain development aim to determine biological and environmental factors that impact developmental trajectories (Deoni et al., 2012; Lebel & Deoni, 2018; Seiberlich et al., 2020). Notable initiatives such as the Human Connectome Project Development (HCP-D) (Somerville et al., 2018), Baby Connectome (BCP) (Howell et al., 2019) and Adolescent Brain Cognitive Development (ABCD) (Casey et al., 2018) projects have included $T_1$-weighted ($T_1$w), $T_2$w, functional and diffusion imaging in their MRI protocols to assess brain morphometry, as well as functional and structural connectivity during neurodevelopment. In addition to these measures, the more recent Healthy Brain and Child Development (HBCD) project aims to use quantitative MRI (qMRI) to estimate $T_1$ and $T_2$ maps (Dean et al., 2024; National Institute on Drug Abuse, 2021). This is beneficial because qMRI yields quantitative parameter maps in physical units (e.g., $T_1$ and $T_2$) that are biophysically interpretable, offering both increased sensitivity and improved specificity to variations in tissue microstructure and composition compared with conventional contrast-weighted imaging, where signal reflects a complex combination of multiple tissue properties, and qMRI enables partial disentanglement of these contributions (Weiskopf et al., 2013; Carter et al., 2025; Saltarelli et al., 2025). It allows for sensitive assessment of subtle developmental changes, identification of physiological alterations undetected by qualitative imaging, and demonstrates higher inter-site reproducibility compared to contrast-weighted imaging (Seiberlich et al., 2020; Tofts, 2005).

However, qMRI sequences are often encoding intensive and may require prohibitively long acquisition times. This is particularly relevant when scanning children, who often have difficulty staying still during prolonged MRI exams. This is the reason why ~25% of all pediatric clinical MR exams are performed under sedation or general anesthesia, and the rate is even higher in children aged 1-7 years (Afacan et al., 2016). Sedation cannot be used to mitigate motion in research studies, since it can only be performed when clinically indicated. Even when scanning adults, the prevalence of motion artifacts is a significant issue, with an estimated 20% of scans requiring repetition due to motion, resulting in substantial additional costs (Andre et al., 2015).

Further, qMRI faces significant technical challenges including cross-vendor availability, which is of concern in multi-site, multi-vendor studies such as the HBCD. Owing to its availability on major scanner vendors, 3D-quantification using an interleaved Look-Locker acquisition sequence with $T_2$ preparation pulse (3D-QALAS) (Dean et al., 2024; Fujita et al., 2024; Kvernby et al., 2014) was chosen as the qMRI acquisition sequence in HBCD. Although other emerging qMRI techniques yield valuable relaxometry information, their multi-vendor availability is limited; e.g. MR Multitasking (Christodoulou et al., 2018) and echo planar time resolved imaging (Wang et al., 2022) have been developed as research sequences on Siemens whereas MR Fingerprinting is a Siemens product with availability as a research sequence on GE (Cao et al., 2022; Gómez et al., 2020; Jiang et al., 2017), with OpenMRF emerging as a new open-source solution for vendor-neutral MRF implementation (Griesler et al., 2026). While 3D-QALAS can be disseminated to the HBCD study sites using all three vendors, its vendor-native implementations lack harmonization because of differences in sequence timing, gradient and radiofrequency (RF) waveforms as well as reconstruction algorithms. These differences hamper cross- vendor reproducibility (A G Teixeira et al., 2020; Boudreau et al., 2024; Karakuzu et al., 2022).

The development of biomarkers to accurately characterize early brain maturation is crucial for understanding neurodevelopment and identifying emerging abnormalities. The myelin sheath is an important marker of neuro-development and a target in many metabolic and neurodegenerative disorders (Nave, 2010; Paus et al., 2008). Assessing myelin development is essential for understanding brain maturation and potential deviations from expected trajectories. Although qMRI assessment of myelin has been explored for over two decades, current techniques suffer from long scan times, poor resolution, sensitivity to motion and variable accuracy (Lee et al., 2021).

Taken together, qMRI has the potential to provide unique and quantitative information sensitive to normal developmental changes and early abnormalities. However, current methods lack multi-vendor availability and harmonization, fail to provide motion-robust and efficient scans suitable for pediatric imaging, and do not offer reliable myelin sensitivity alongside $T_1$ and $T_2$ relaxometry maps.

We propose a vendor-agnostic qMRI sequence capable of estimating whole-brain relaxometry and MWF maps from an efficient, motion-compensated 4.7 min acquisition at 1 mm isotropic resolution at 3T. We complement this sequence with regularized temporal subspace reconstruction algorithms including retrospective motion correction and a rapid calibration scan for flip angle ($B_1^+$) and coil sensitivity ($B_1^-$) estimation. We also propose a Bloch-simulation-based dictionary matching algorithm for unified parameter estimation using the reconstructed volumes, thus further boosting reproducibility. The technique is validated through simulations, in system phantom and in volunteers, and finally demonstrated in pediatric subjects. The main contributions of our work are as follows:

- Vendor-agnostic, open-source qMRI pulse sequence with joint and efficient relaxometry and MWF mapping.
- Rapid and harmonized calibration scan for $B_1^+$ and $B_1^-$ mapping.

- Motion-compensation by self-navigation and a unified image reconstruction pipeline, including retrospective motion-corrected subspace reconstruction and parameter estimation.
- Validation in simulation, system phantom, adults, and in pediatric cases.
- All pulse sequences and processing code are provided under an open-source license.

## 2. Methods

### 2.1 Myelin-sensitive sequence design and processing overview

The typical 3D-QALAS sequence incorporates five fast low angle shot (FLASH) readouts (Frahm et al., 1986; Haase et al., 1986), interleaved with a $T_2$-preparation module and an adiabatic inversion pulse. The $T_2$-preparation module uses four hyperbolic secant adiabatic refocusing pulses and two standard hard 90-degree pulses for tip-down/up (Jenista et al., 2013), with a $TE_{T2prep}$ of 100 ms. Due to the short $T_2$ of myelin water (typically less than 25 ms at 3T) (Labadie et al., 2014; Laule et al., 2007; Lee et al., 2021), its signal decays rapidly, and such $T_2$-preparation modules with longer echo times lack adequate sensitivity to this component.

To address the limitations of standard 3D-QALAS (Fujita et al., 2019; Kvernby et al., 2014) in detecting short $T_2$ components, we designed a modified acquisition scheme termed MWF-QALAS, where we added an additional short $T_2$-preparation pulse (TE = 20 ms) followed by an extra readout, resulting in six total acquisitions per TR (**Figure 1A**). The sampling at multiple effective inversion times occurs within the same TR through segmented Look-Locker-style sampling, without requiring multiple TRs with different prescribed inversion times. This modification results in two $T_2$-preparation modules, each with its own readout: the first (TE=20ms, orange block) targeting myelin water, and the second (TE=80ms, blue block) sensitive to longer $T_2$ species as in the original 3D-QALAS. After reconstructing the MWF-QALAS signal evolution using regularized retrospective motion-corrected temporal subspace reconstruction (section 2.2), we use Bloch equation based dictionary matching (**Figure 1B**) for relaxometry parameter estimation (Cho et al., 2024) and a partial volume dictionary (**Figure 1C**) to estimate MWF (Chen et al., 2019; Deshmane et al., 2019).

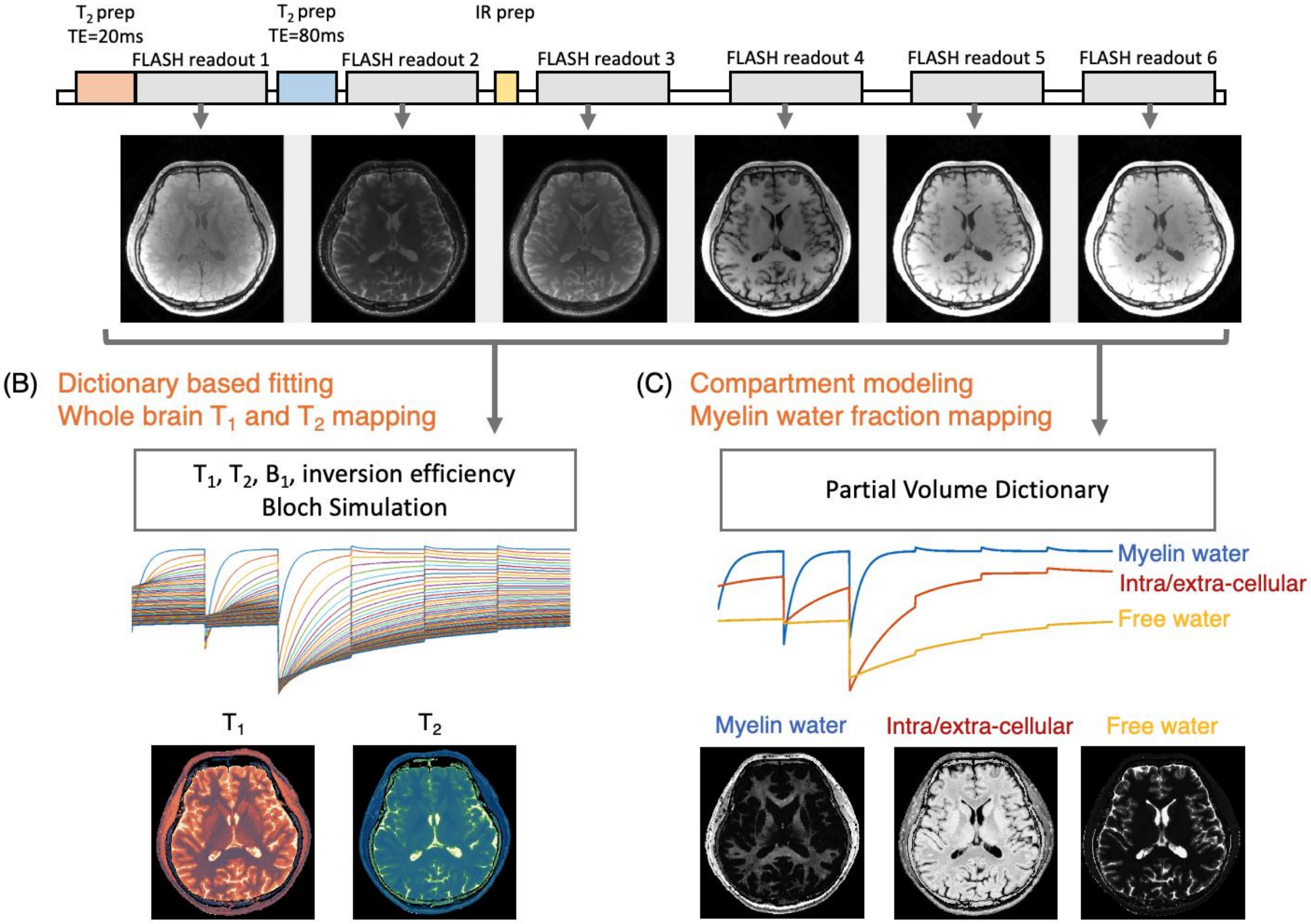

**Figure 1.** Overview of the MWF-QALAS sequence design and processing pipeline. (A) Sequence design. Two $T_2$-preparation pulses (TE = 20 ms and TE = 80 ms) are inserted to improve sensitivity to myelin water, while an inversion pulse is also used for $T_1$ sensitization. Six FLASH readouts are acquired to capture the dynamic signal evolution (illustrated by the images below the pulse sequence). (B) Dictionary-based $T_1$ and $T_2$ mapping. A set of Bloch-simulated signal evolutions (incorporating $T_1$, $T_2$, $B_1$, and inversion-pulse efficiency) is generated to form a "dictionary." Each voxel's measured signal response is matched to the best-fitting entry in this dictionary, yielding whole-brain $T_1$ and $T_2$ maps. (C) Myelin water fraction (MWF) quantification. A multi-compartment model partitions the voxel signal into myelin water, intra/extra-cellular water, and free water components. A partial-volume dictionary is used to solve for each compartment's contribution, providing voxel-wise MWF maps. Example simulated signal evolutions for each water component demonstrate their distinct decay characteristics.

## 2.2 Unified image reconstruction pipeline including retrospective motion correction

Raw k-space data is reconstructed offline in MATLAB with a unified pipeline for different vendors and scanners. The proposed pipeline of motion estimation and reconstruction framework is shown in **Figure 2**. Data were acquired with a 3D Cartesian trajectory using a variable-density, center-out spiral profile ordering in the phase-encoding (ky−kz) plane. This ordering preserves a Cartesian trajectory while sampling central k-space more densely. The sampling involves a center-out acquisition with each echo train designed to sample complementary frequencies across contrasts and across TRs. For each TR, FLASH readouts #4 to #6 are aggregated to obtain the k-space sample needed for motion estimation (right column). These blocks exhibit similar contrast (**Figure 1**), which enables their data to be aggregated to collect more k-space information within a single TR. Using multiple contrast-consistent readouts increases the effective data available for motion estimation while avoiding inconsistencies that would arise from mixing blocks with substantially differing contrast. The k-space center signal (DC signal) with a radius of 3 was consistently sampled per TR to maintain stable contrast navigation images.

This undersampled k-space data per TR is used to reconstruct volume data using SENSE (Pruessmann et al., 1999) and estimate motion state at each TR. The time-resolved navigation images had matrix size 32x32x32 with an effective acceleration factor of ~18. These time-resolved data are used to estimate rigid 3D head motion by calculating the six-degree-of-freedom transformation $T_t$ for each volume relative to a reference volume.

This transformation $T_t$ is incorporated into the retrospective motion-corrected temporal subspace reconstruction. For each TR, translations were applied as phase shifts in k-space, and rotations were applied by rotating the k-space coordinates, following the standard retrospective k-space motion-correction framework (Gallichan et al., 2016). Because the rotated k-space samples are no longer on the Cartesian grid, the final reconstruction is performed using a NUFFT operator. The MWF-QALAS signal evolution is represented using a low-dimensional temporal subspace, similar to prior multicontrast MRI subspace reconstruction approaches (Tamir et al., 2017; Cao et al., 2022), and the subspace coefficient maps are estimated by solving:

$$\{\hat{c}_l\}_{l=1}^{L} = \underset{\{c_l\}_{l=1}^{L}}{\arg\min} \sum_t \left\| P_t \mathcal{F} C \left( \sum_{l=1}^{L} c_l\,(\mathbf{r}) \phi_l(t) \right) - k_t \right\|_2^2 + \lambda R\left(\{c_l\}_{l=1}^{L}\right).$$

Here, $c_l(r)$ represents the $l$-th subspace coefficient map, $\phi_l(t)$ represents the $l$-th temporal basis function, $L$ is the number of temporal basis functions, $k_t$ represents the acquired multicoil k-space data at time point $t$, $P_t$ represents the sampling operator, $C$ represents the coil-sensitivity profiles, and $F$ represents the motion-aware NUFFT operator incorporating the transformation. $R$ is the wavelet regularization term with regularization parameter $\lambda$. Regularized temporal subspace reconstruction was performed using the Berkeley Advanced Reconstruction Toolbox (BART) (Uecker et al., 2015).

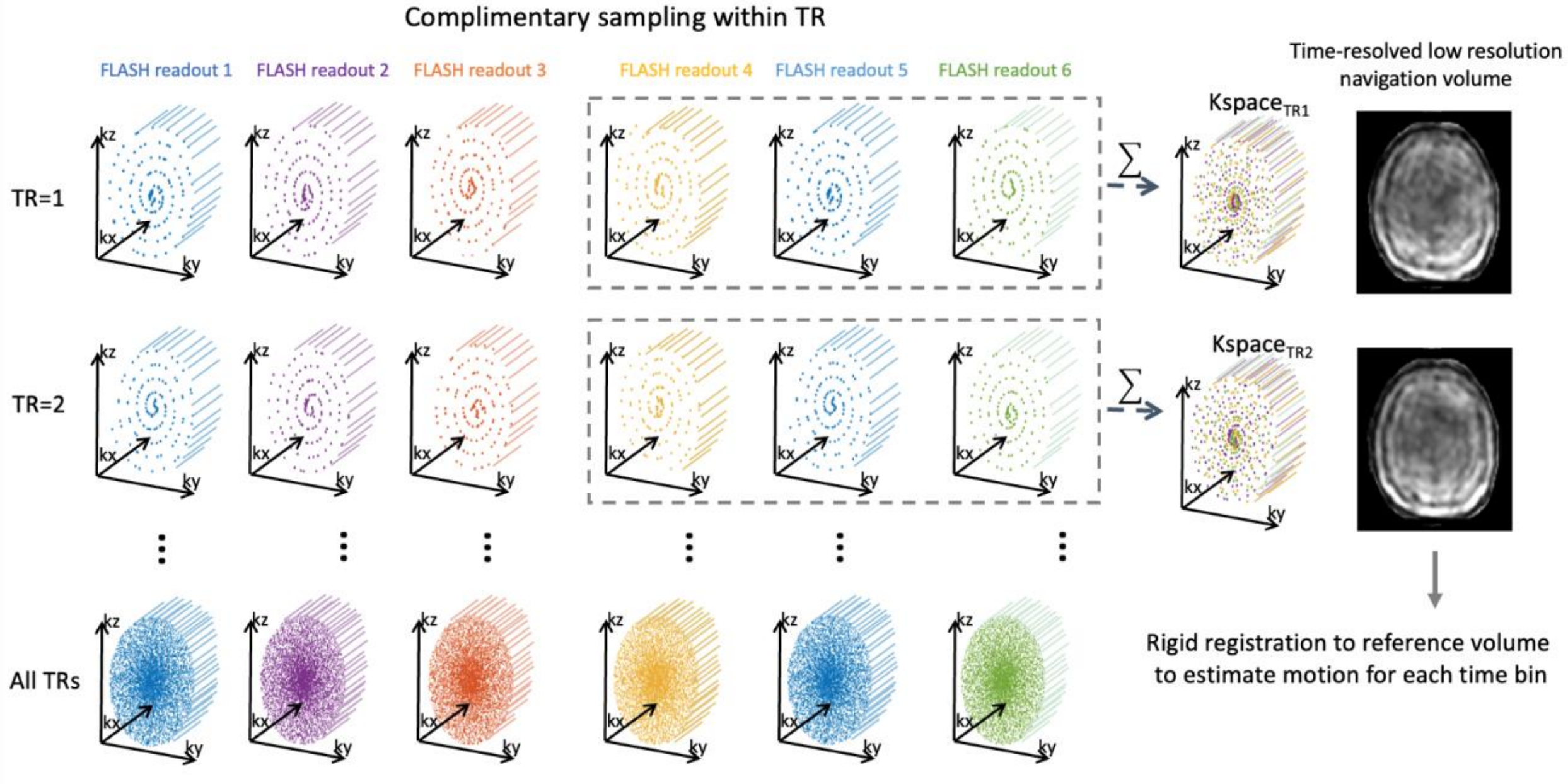

**Figure 2.** Unified image reconstruction pipeline with retrospective motion correction. (A) Acquisition was based on 3D Cartesian trajectory with spiral profile ordering (center-out ky–kz). The sampling involves a center-out acquisition with each echo train designed to sample complementary frequencies across contrasts and across TRs. For each TR, FLASH readouts no. 4 to 6 are aggregated to obtain k-space to be used for motion estimation (right column). This undersampled k-space data per TR is used to reconstruct low-resolution navigation volumes using SENSE and estimate motion state at each TR. The transformation matrix $T_t$ at each TR is subsequently incorporated into the retrospective motion-corrected temporal subspace reconstruction, where translations are applied as k-space phase shifts and rotations are applied by rotating the k-space coordinates before NUFFT-based reconstruction.

### 2.3 Joint parameter estimation with multi-compartment dictionary fitting

To estimate tissue relaxation parameters, the full MWF-QALAS temporal signal evolution was reconstructed from the estimated subspace coefficients and matched to voxel-wise Bloch-simulated signal evolutions using a precomputed dictionary (Cho et al., 2024). $B_1^+$ correction was applied using maps derived from the turbo Actual Flip angle Imaging (tAFI) scan (section 2.4), along with an estimation of inversion efficiency. The same inversion pulse, with identical waveform and timing, was used across vendors in our implementation. Therefore, variations in inversion efficiency primarily arise from differences in $B_1^+$ and off-resonance conditions, whereas vendor-provided inversion pulses may introduce additional variability because of differences in pulse design and implementation. The QALAS images were rigidly registered to tAFI volume before fitting using FMRIB Software Library. The step size and number of dictionary entries were empirically selected to balance parameter granularity with computational feasibility. $T_1$ values were sampled with a 10-ms step size up to 3000 ms, and with a 100-ms step size up to 5000 ms. $T_2$ values were sampled with a 2-ms step size up to 350 ms, and with a 20-ms step size up to 500 ms. B1+ values were sampled with a 0.05 step size between 0.65 and 1.35.

For myelin water fraction estimation, we employed a three-compartment model (Chen et al., 2019; Deshmane et al., 2019; Labadie et al., 2014) consisting of: (i) myelin water ($T_1$ = 150 ms, $T_2$ = 15 ms), (ii) intra-/extra-cellular water with two subcomponents ($T_1$ = 1100 ms, $T_2$ = 50 ms and $T_1$ = 1300 ms, $T_2$ = 70 ms), and (iii) free water ($T_1$ = 4500 ms, $T_2$ = 500 ms). The relative signal contributions of each compartment were estimated through multi-compartment dictionary fitting, providing voxel-wise MWF maps.

### 2.4 Harmonized $B_1^+$ and $B_1^-$ calibration scan

Incorporating a flip angle map into the parameter estimation step of qMRI techniques mitigates biases (Boudreau et al., 2017). Thus, in addition to the MWF-QALAS sequence, a rapid tAFI technique was proposed by extending the original AFI sequence (Yarnykh, 2007) to harmonize the various calibration steps in addition to the imaging sequence itself.

The proposed sequence diagram is shown in **Supplementary Figure 1**. Original AFI consists of two identical RF pulses followed by two delay periods, $TR_1$ and $TR_2$. Gradient echo (GRE) signals are acquired after each RF pulse. It has been demonstrated that if the transverse magnetization is completely spoiled, the ratio of signal intensities ($r = S_2/S_1$) depends on the flip angle of applied pulses and is highly insensitive to $T_1$ (Yarnykh, 2007). An analytical relation between the flip angle and the ratio of the two GRE signals can be exploited to estimate the $B_1^+$ map as follows:

$$\theta \approx \arccos[(rn - 1)/(n - r)]$$

Where r is the ratio of the two GRE signals ($S_2/S_1$), n is the ratio of the two TRs ($TR_2/TR_1$), and $\theta$ is the estimated flip angle. The need for large spoiler gradients and $TR_2 >> TR_1$ constraints create dead time, which we exploited by inserting additional GRE readouts to create a faster, “turbo” version of the sequence. The same mapping acquisition was used for coil sensitivity estimation from the fully sampled 24x24 ACS region. By sampling four echoes in each of the two TRs and by R=4-fold undersampling in the peripheral (non-ACS) k-space, an acquisition at 4mm isotropic resolution required 31 seconds.

**2.5 Numerical Simulation**

A numerical $T_1$-$T_2$ grid phantom was prepared, consisting of a range of predefined $T_1$ and $T_2$ relaxation times arranged in a systematic grid. This phantom served as the reference for validating the sequence's performance. Using the signal model of each of the imaging sequences, source contrast images were simulated. To mimic real-world experimental conditions, 1%-5% Gaussian noise was added to the simulated images to introduce variability akin to noise in actual measurements.

Subsequently, $T_1$ and $T_2$ relaxation maps were computed from the simulated images using a dictionary-matching approach. The dictionary-matching technique relied on a precomputed library of signal intensities corresponding to different combinations of $T_1$ and $T_2$ values (Cho et al., 2024), allowing precise estimation of these parameters from the noisy data. The obtained $T_1$ and $T_2$ maps were then quantitatively compared against the numerical reference phantom. The comparison involved calculating the percent difference and absolute difference between the derived maps and the reference values in the numerical phantom, providing a measure of the accuracy of the sequence for a range of $T_1$ and $T_2$ combinations.

**2.6 Data acquisition**

The FOV parameters for MWF-QALAS and tAFI were tailored to align with HBCD-compliant pediatric (Dean et al., 2024) and ADNI-compliant adult (Weiner et al., 2017) protocols. For MWF-QALAS, the shared parameters included sagittal orientation; TR = 4500 ms; TE = 2.29 ms; TI = 110/1010/1910/2810/3710 ms; echo spacing = 5.8 ms; flip angle = 4°; echo train length = 128; and acceleration factor (R) = 5. Protocol-specific parameters were as follows:

- **HBCD-compliant:** bandwidth = 326 Hz/pixel; field of view (FOV) = 228 × 228 × 176 $mm^3$; matrix = 228 × 228 × 176; acquisition time (TA) = 4 min 41 sec.
- **ADNI-compliant:** bandwidth = 279 Hz/pixel; FOV = 256 × 240 × 208 $mm^3$; matrix = 256 × 240 × 208; TA = 5 min 46 sec.

For tAFI, the acquisition parameters were consistent across protocols for $TR_1$ = 19 ms; $TR_2$ = 95 ms; TE = 2.5 ms; echo spacing = 3.5 ms; flip angle = 60°; echo train length = 4. Protocol-specific parameters were:

- **HBCD-compliant:** bandwidth = 330 Hz/pixel; FOV = 228 × 228 × 176 $mm^3$; matrix = 56 × 56 × 44; TA = 31 sec.
- **ADNI-compliant:** bandwidth = 325 Hz/pixel; FOV = 256 × 240 × 208 $mm^3$; matrix = 64 × 60 × 52; TA = 33 sec.

Both the MWF-QALAS and the tAFI sequences were implemented using Pulseq v1.4.1(Layton et al., 2017)(Nielsen & Noll, 2018) in MATLAB, and executed on the following 3-T scanners:

- Site A, MAGNETOM Prisma (Siemens Healthineers, Forchheim, Germany)
- Site A, MAGNETOM Skyra (Siemens Healthineers, Forchheim, Germany)
- Site B, MAGNETOM Prisma (Siemens Healthineers, Forchheim, Germany)
- Site C, SIGNA Premier XT (GE HealthCare, Waukesha, WI)

The third platform (Site B, Prisma), corresponding to a clinical pediatric site, was used exclusively for pediatric scans in this study.

### 2.7 System phantom validation

A NIST/ISMRM phantom was scanned with a spin-echo (SE) scan with multiple inversion recovery (IR) times for $T_1$ mapping, and multiple single-echo SE for $T_2$ mapping for references, as shown in **Supplementary Table 1**.

To measure relaxation times in the phantom data, circular spherical regions of interest (ROIs) with a diameter of 10 mm were placed in the phantom's spheres with clinically relevant relaxation times, covering the typical physiological ranges of $T_1$ and $T_2$ values for human tissues (approximate ranges: 200–2500 ms for $T_1$ and 30–300 ms for $T_2$).

### 2.8 In vivo validation

This prospective study was approved by the Institutional Review Board and conducted in accordance with HIPAA regulations. Three healthy volunteers were scanned at multiple institutions using scanners from different vendors (2 male, 1 female; age range, 25–35 years). No sedation was used. Written informed consent was obtained from all healthy volunteers. Separately, pediatric patients undergoing brain MRI for clinical indications were recruited after obtaining consent from their guardians. In vivo validation for MWF mapping was performed using multi-echo 3D-GRASE research application (Piredda et al., 2021). The scanning parameters of 3D-GRASE (1.6 mm isotropic voxels) were as follows: Sagittal orientation; TR = 1040 ms; TE = 11.28 to 360.96 ms (32 contrasts in total); echo spacing = 1.31 ms; flip angle = 185°; echo train length = 128; bandwidth = 1240 Hz/pixel; FOV = 228 × 200 × 154 $mm^3$; matrix = 144 × 124 × 96; TA = 10 min 4 sec. To assess clinical feasibility, pediatric subjects were included in the validation cohort.

MWF maps were segmented into anatomical structures as listed in **Supplementary Table 2** using FastSurfer (Henschel et al., 2020), and structure-wise mean values were extracted for subsequent comparison between MWF-QALAS and 3D-GRASE and across scanner platforms.

### 2.9 Motion correction validation

To evaluate the robustness of the proposed motion-corrected reconstruction, both prospective motion experiments and retrospective motion simulations were performed. Three prospective acquisitions were obtained under different motion conditions: Experiment 1, no intentional motion; Experiment 2, continuous head motion throughout the scan; and Experiment 3, stepwise head repositioning approximately every 1 min. Motion trajectories were estimated from the self-navigation images and used for retrospective motion correction.

To evaluate reconstruction accuracy in the presence of motion, retrospective motion simulations were performed using the motion-free acquisition as the reference dataset. Because a true reference map is not available during an actual scan with motion, the motion-free reconstruction was treated as the reference and used to generate controlled motion-corrupted datasets.

- In Experiment 4, the motion trajectory estimated from Experiment 2 was applied to the motion-free data, and contrasts were reconstructed under three conditions: without motion correction, with motion correction using the estimated trajectory, and with motion correction using the known applied trajectory.
- In Experiment 5, the same reference dataset was used while the applied motion trajectory was scaled by factors of 0.1, 0.5, 1, and 2 to assess the effect of motion magnitude on reconstruction accuracy.
- In Experiment 6, a randomly varying stepwise motion trajectory was applied, updated every 1 min, with a maximum rotation of 1° and a maximum translation of 3 mm.

### 2.10 Statistical analysis

All relaxometry model fittings were performed voxel-wise using the full signal evolution at each voxel to generate quantitative parametric maps (e.g., $T_1$ and $T_2$). ROI analyses were then conducted on the derived maps by extracting voxel-wise parameter values within each predefined region. ROI-level metrics were computed as the mean of the fitted voxel-wise estimates. Signal averaging within ROIs prior to model fitting was not performed, in order to avoid bias introduced by nonlinear fitting of averaged signals.

For the system phantom experiments, within-ROI variability was assessed using the coefficient of variation (CV), defined as the standard deviation of voxel-wise measurements within each ROI divided by the corresponding mean ROI value. Linear regression was used to evaluate agreement between QALAS and MWF-QALAS-derived $T_1$ and $T_2$ measurements and the corresponding reference values.

For the in vivo validation, agreement between MWF-QALAS-derived and 3D-GRASE-derived MWF measurements was evaluated using region-wise measurements obtained across multiple brain structures in three healthy volunteers. Pearson's correlation coefficient was calculated to quantify linear association. Agreement and reliability were assessed using the intraclass correlation coefficient, and Bland–Altman analysis was performed to evaluate systematic bias and limits of agreement between MWF-QALAS- and 3D-GRASE-derived MWF measurements. Cross-vendor agreement in $T_1$, $T_2$, and MWF measurements was assessed using ICC(2,1), based on a two-way random-effects, single-measure agreement model (Shrout & Fleiss, 1979). Pairwise Pearson correlation coefficients and Bland–Altman analyses were additionally performed to evaluate linear association, systematic bias, and limits of agreement between scanner pairs.

For the retrospective motion simulations, reconstruction accuracy was quantified relative to the motion-free reference using normalized root-mean-square error (NRMSE) across all six contrasts.

All statistical analyses were performed using R version 4.5.3 (R Foundation for Statistical Computing, Vienna, Austria). A two-sided P value < 0.05 was considered statistically significant.

## 3. Results

### 3.1 Numerical Simulation

**Figure 3** demonstrates the accuracy of $T_1$ and $T_2$ estimates under different $TE_{T2prep}$ conditions when compared to the reference maps. For $T_1$ estimation, both $TE_{T2prep}$=100 ms (standard QALAS) and $TE_{T2prep}$=20 ms & 80 ms (proposed MWF-QALAS) produced maps that closely matched the reference $T_1$, with minor relative differences in small $T_1$ values. For $T_2$ estimation, while the single $TE_{T2prep}$=100 ms method provided reasonable $T_2$ estimations, it exhibited slightly larger deviations, particularly for lower $T_2$ values. In contrast, the proposed $TE_{T2prep}$ = 20 ms & 80 ms approach achieved improved fidelity, particularly in capturing short $T_2$ values, as seen in the relative difference maps. The mean $T_1$ bias across the full range was –2.2 ms for 3D-QALAS and –1.3 ms for MWF-QALAS. For $T_2$ estimates, the mean bias across all values was 3.0 ms and 2.0 ms, respectively. In the short $T_2$ range (<30 ms), the bias was reduced from -3.1 ms with 3D-QALAS to -0.1 ms with MWF-QALAS. These findings suggest that the addition of the shorter $TE_{T2prep}$ times improves the accuracy of $T_2$ mapping for a wide range of $T_2$ values, while maintaining comparable performance for $T_1$ estimation. In the proposed $TE_{T2prep}$ = 20 ms & 80 ms approach, noticeable deviations occurred only for extremely short $T_1$ values below 10 ms, a range in which the signal is largely insensitive to further changes in $T_1$. Within the physiologic range relevant to brain tissue, where $T_1$ values are greater than roughly 100 to 150 milliseconds, the relative differences remained less than 2%, demonstrating robust parameter estimation across the simulated tissue properties. Additional analyses at noise levels of 1%, 3%, and 5% are presented in **Supplementary Figure 2**, with the corresponding absolute-error statistics reported in **Supplementary Tables 3 and 4.**

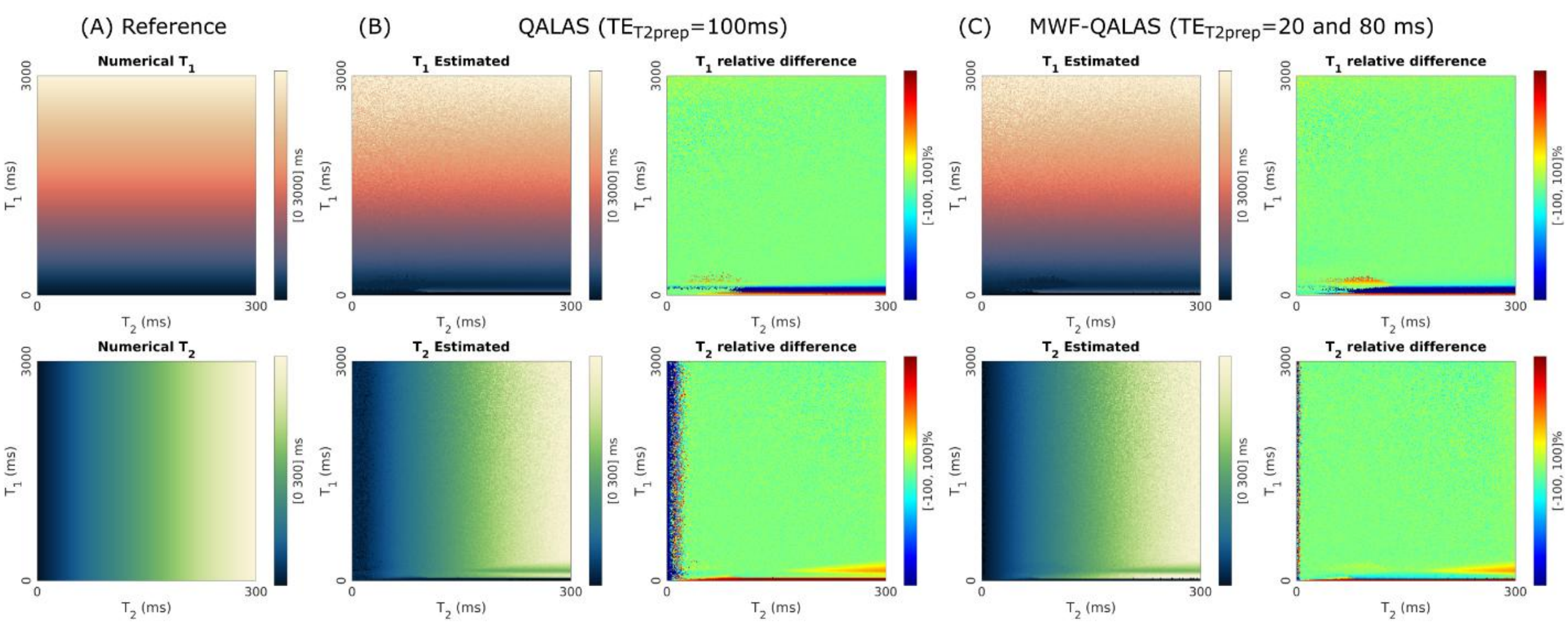

**Figure 3.** Numerical simulation of sequence mapping accuracy. (A) Reference $T_1$ (top)- $T_2$ (bottom) grid used as ground truth for the numerical simulation. (B) QALAS results with a single $T_2$-preparation time of 100 ms: the left panels show the estimated $T_1$ and $T_2$ maps, while the right panels display the corresponding relative-difference maps (in percent) compared to the reference. Underestimation is evident in short $T_2$ values, regardless of the $T_1$ value. (C) MWF-QALAS results incorporating two $T_2$-prep times (20 ms and 80 ms): the left panels show the estimated $T_1$ and $T_2$ maps, and the right panels are the relative-difference maps. Notably, the MWF-QALAS reduces estimation bias in tissues with shorter $T_2$ values relative to using a single 100 ms $T_2$-prep.

### 3.2 Phantom validation

We evaluated the mapping accuracy of MWF-QALAS using a NIST/ISMRM system phantom with known $T_1$ and $T_2$ relaxation values. **Figure 4** compares the quantitative maps obtained from reference techniques (inversion recovery for $T_1$, spin echo for $T_2$), standard QALAS, and MWF-QALAS. Both QALAS-based methods showed good agreement with the reference, but MWF-QALAS demonstrated notably improved performance for short-$T_2$ compartments (see the inner vials). Compared with standard QALAS, MWF-QALAS exhibited lower within-ROI variability. This difference was reflected in the ROI coefficient of variation, defined as the standard deviation of voxel values within the ROI divided by the mean ROI value, which decreased from 26.7 ± 20.8% with standard QALAS to 13.1 ± 10.8% with MWF-QALAS across all vials, and from 33.4 ± 21.3% to 15.0 ± 12.4% in the low $T_2$ vials ($T_2$ < 100 ms). In the low-$T_2$ vials, standard QALAS also showed a greater negative bias of −35.4 ± 26.8%, whereas MWF-QALAS showed a smaller positive bias of 12.1 ± 22.2%. Quantitative comparison via linear regression (**Figure 4B**) showed slopes closer to unity and improved linearity with MWF-QALAS across the full fitted ranges for both $T_1$ (slope: 1.18 for QALAS vs. 1.13 for MWF-QALAS; $R^2$ = 0.9971 vs. 0.9984) and $T_2$ (slope: 0.77 vs. 0.86; $R^2$ = 0.9803 vs. 0.9932). In the short-$T_2$ range ($T_2$ ≤ 70 ms), slope bias was reduced from 0.78 with QALAS to 0.91 with MWF-QALAS, and linearity improved from $R^2$ = 0.7929 to $R^2$ = 0.9559. See **Supplementary Figure 3** for log scale plot.

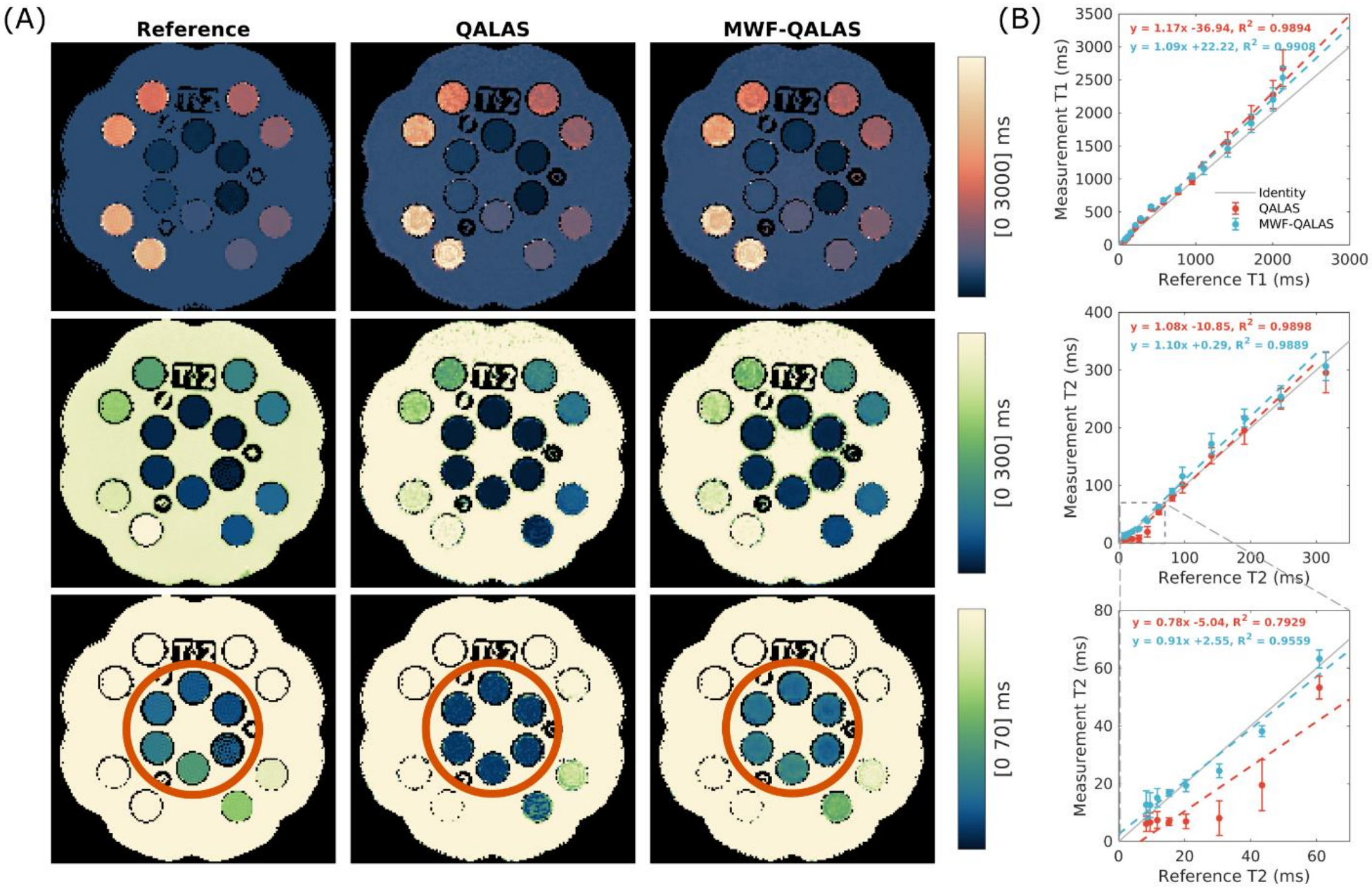

**Figure 4. $T_1$ and $T_2$ Mapping Validation in NIST/ISMRM Phantom.** (A) Quantitative $T_1$ and $T_2$ maps obtained with reference methods (inversion recovery for $T_1$ and spin echo for $T_2$), QALAS, and MWF-QALAS. Lower-range $T_2$ scale display shorter $T_2$ values relevant for myelin (bottom row). The orange circle in the short-$T_2$ row highlights vials with short $T_2$ values. (B) Scatter plots of measured versus reference $T_1$ (top) and $T_2$ (middle) values for the QALAS (red) and MWF-QALAS (blue) methods, with linear regression fits calculated over $T_1$ values from 0 to 3000 ms and $T_2$ values from 0 to 300 ms. The bottom panel shows an enlarged view of the short-$T_2$ range ($T_2$ ≤ 70 ms), with separate linear regression fits calculated using only the short-$T_2$ vials. The corresponding regression equations and $R^2$ values are shown in each panel. The error bars represent the standard deviation within each vial. MWF-QALAS demonstrates improved stability, particularly for shorter-$T_2$ samples.

### 3.3 In vivo validation of myelin water fraction

To validate the MWF estimations, we applied MWF-QALAS to three healthy adult subjects and compared the obtained maps to those from a reference multi-echo 3D-GRASE research application which required 10:04 min (Piredda et al., 2021). **Figure 5** displays the resulting in vivo $T_1$, $T_2$, and tissue compartment maps using MAGNETOM Prisma (Siemens Healthineers, Forchheim, Germany). **Supplementary Figure 4** presents a side-by-side comparison across multiple views, including zoomed-in regions and voxel-wise absolute-difference maps, demonstrating the preservation of fine anatomical details at 1.0-mm isotropic resolution compared to the more commonly used 1.6-mm isotropic resolution. The MWF maps from MWF-QALAS showed close correspondence with those from 3D-GRASE, particularly in expected high-MWF regions such as the corpus callosum and internal capsule. Quantitative comparison of structure-wise MWF measurements across three healthy adult subjects yielded an intraclass correlation coefficient of 0.732 (95% CI: 0.613–0.819), Pearson's correlation coefficient of 0.757 (95% CI: 0.645–0.837), and a Bland–Altman mean difference of -0.0074 with limits of agreement from -0.0544 to 0.0396 (**Supplementary Figures 5 and 6**). The voxel-count-weighted mean difference across segmented brain regions was 0.0164, corresponding to a voxel-count-weighted mean percent difference of 13.02%.

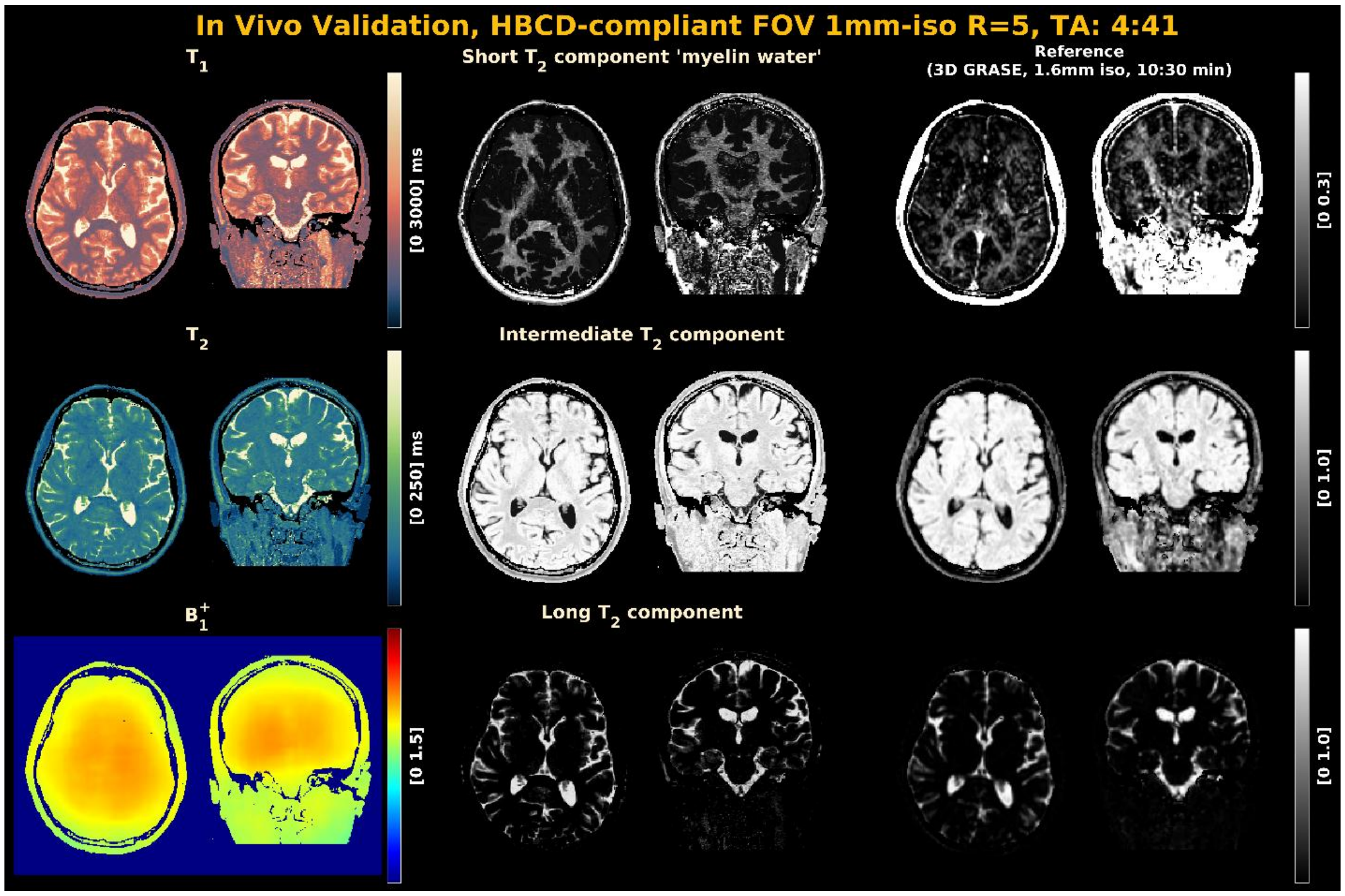

**Figure 5.** In vivo validation of the proposed MWF-QALAS in a healthy adult. The myelin water fraction, corresponding to the short $T_2$ component, is shown alongside $T_1$, $T_2$, intermediate $T_2$ ('intra/extra-cellular water'), and long $T_2$ ('free water') maps. Reference images related to the MWF mapping, in the same subject were obtained with a multi-echo gradient and spin echo (GRASE) sequence at 1.6-mm isotropic resolution with 10:30 minutes total scan time.

### 3.4 In vivo validation across scanners

**Figure 6** presents in vivo imaging results acquired on MAGNETOM Prisma and Skyra (Siemens Healthineers, Forchheim, Germany) and SIGNA Premier XT (GE HealthCare, Waukesha, WI) scanners from the same healthy adult volunteer. Cross-vendor quantitative analysis was performed using measurements from three healthy adult subjects. Quantitative comparison using ICC(2,1) across vendors yielded values of $T_1$: 0.934 (95% CI: 0.890–0.959), $T_2$: 0.987 (95% CI: 0.982–0.991), and MWF: 0.771 (95% CI: 0.685–0.838). Pairwise Pearson correlation coefficients and Bland–Altman analyses are provided in **Supplementary Figure 7** and **Supplementary Table 5**. Results demonstrate reproducibility comparable to within-vendor reproducibility reported in MRF (Wicaksono et al., 2023). This consistency supports the feasibility of applying the sequence in multi-site, multi-vendor studies.

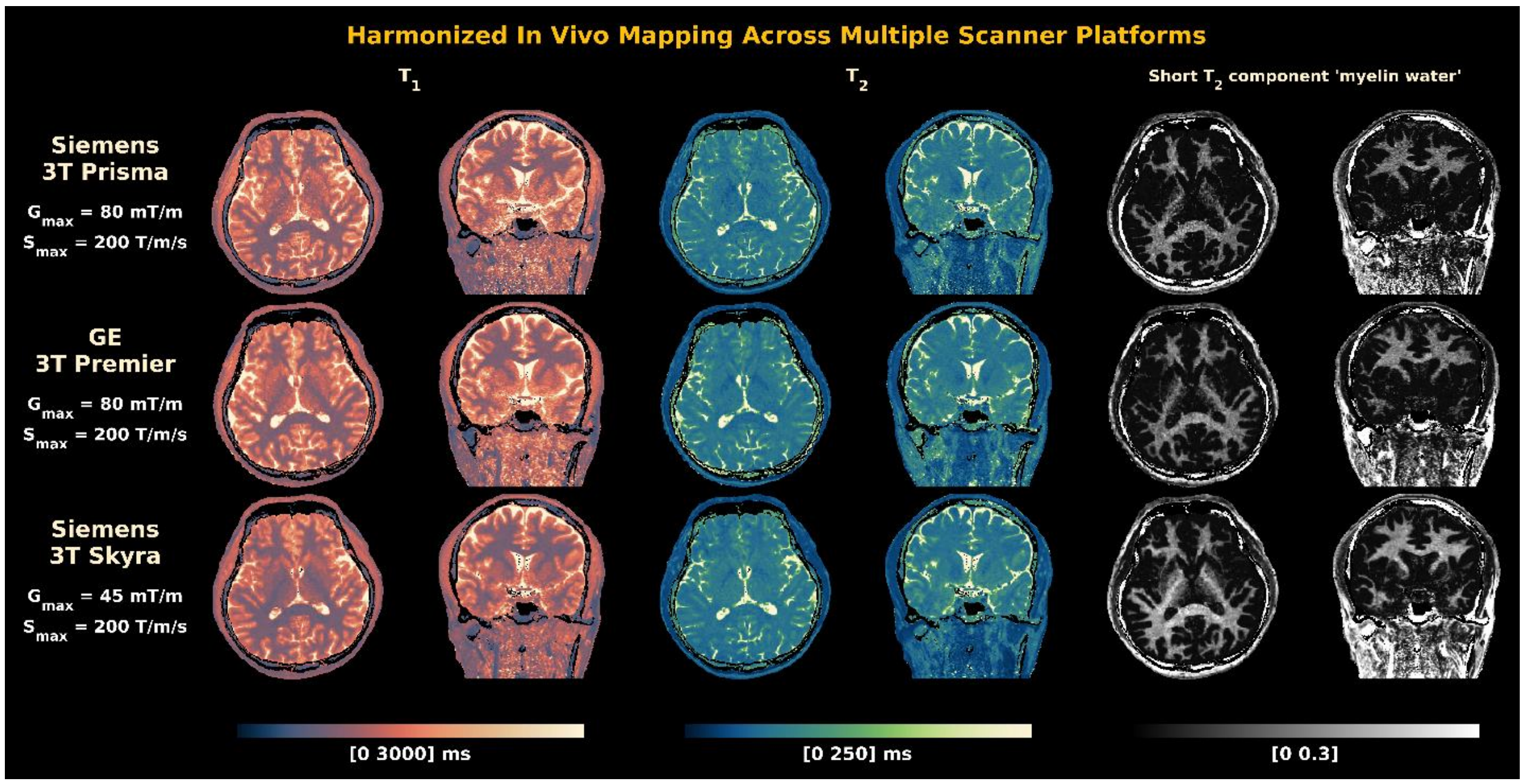

**Figure 6.** Representative maps from different vendor scanners for multisite studies. $T_1$, $T_2$, and myelin water fraction maps were obtained from Siemens and GE scanners. These maps were acquired with a vendor-agnostic implementation, allowing cross-platform compatibility for multisite studies. The acquisition protocol is compatible with varying scanner specifications, where $G_{max}$ represents maximum gradient strength, and $S_{max}$ represents maximum slew rate.

### 3.5 Motion correction validation

In the prospective motion experiments, Experiment 1 was acquired without intentional motion, whereas Experiments 2 and 3 included continuous head motion and stepwise head repositioning, respectively. The estimated motion trajectories showed translations of up to approximately 4 mm and rotations of up to approximately 1° during the continuous-motion experiment, and translations of up to approximately 2 mm and rotations of up to approximately 1° during the stepwise-motion experiment. Compared with reconstructions without motion correction, the motion-corrected reconstructions showed reduced motion-related artifacts and improved preservation of anatomical structure in both motion experiments (**Supplementary Figure 8**).

In the retrospective simulations, Experiment 4 compared reconstructions without motion correction, with correction using the estimated motion trajectory, and with correction using the known applied trajectory. Motion correction reduced the reconstruction error relative to the motion-free reference, with NRMSE values of 29.53%, 13.16%, and 6.21%, respectively (**Supplementary Figure 9**).

Experiment 5 evaluated the robustness of motion correction across increasing motion magnitudes by scaling the trajectory used in Experiment 4 by factors of w=0.1, 0.5, 1, and 2. Motion correction reduced NRMSE from 2.89% to 1.75% at w=0.1, from 16.49% to 5.43% at w=0.5, from 29.53% to 6.21% at w=1, and from 46.64% to 8.04% at w=2. Although residual error increased with motion magnitude, motion correction remained effective across the tested range (**Supplementary Figure 10**).

In Experiment 6, robustness was further evaluated using a randomly varying stepwise motion trajectory updated approximately every 1 min, with maximum rotations of 1° and translations of 3 mm. Motion correction using the estimated trajectory reduced NRMSE from 50.84% to 20.47% and visibly reduced motion-related artifacts, although residual differences remained relative to the motion-free reference (**Supplementary Figure 11**).

### 3.6 Pediatric cases

To illustrate the clinical applicability of the proposed method in pediatric populations, we present a representative case of a normally developing 12-year-old male subject (**Figure 7**). Despite observable head motion during the scan, retrospective motion correction effectively mitigated the associated artifacts, resulting in high-quality $T_1$, $T_2$, and MWF maps. Furthermore, **Figure 8** displays MWF maps from pediatric subjects of varying ages (alongside maps from an adult volunteer, for comparison), demonstrating progressive changes in brain myelination, consistent with expected neurodevelopmental stages.

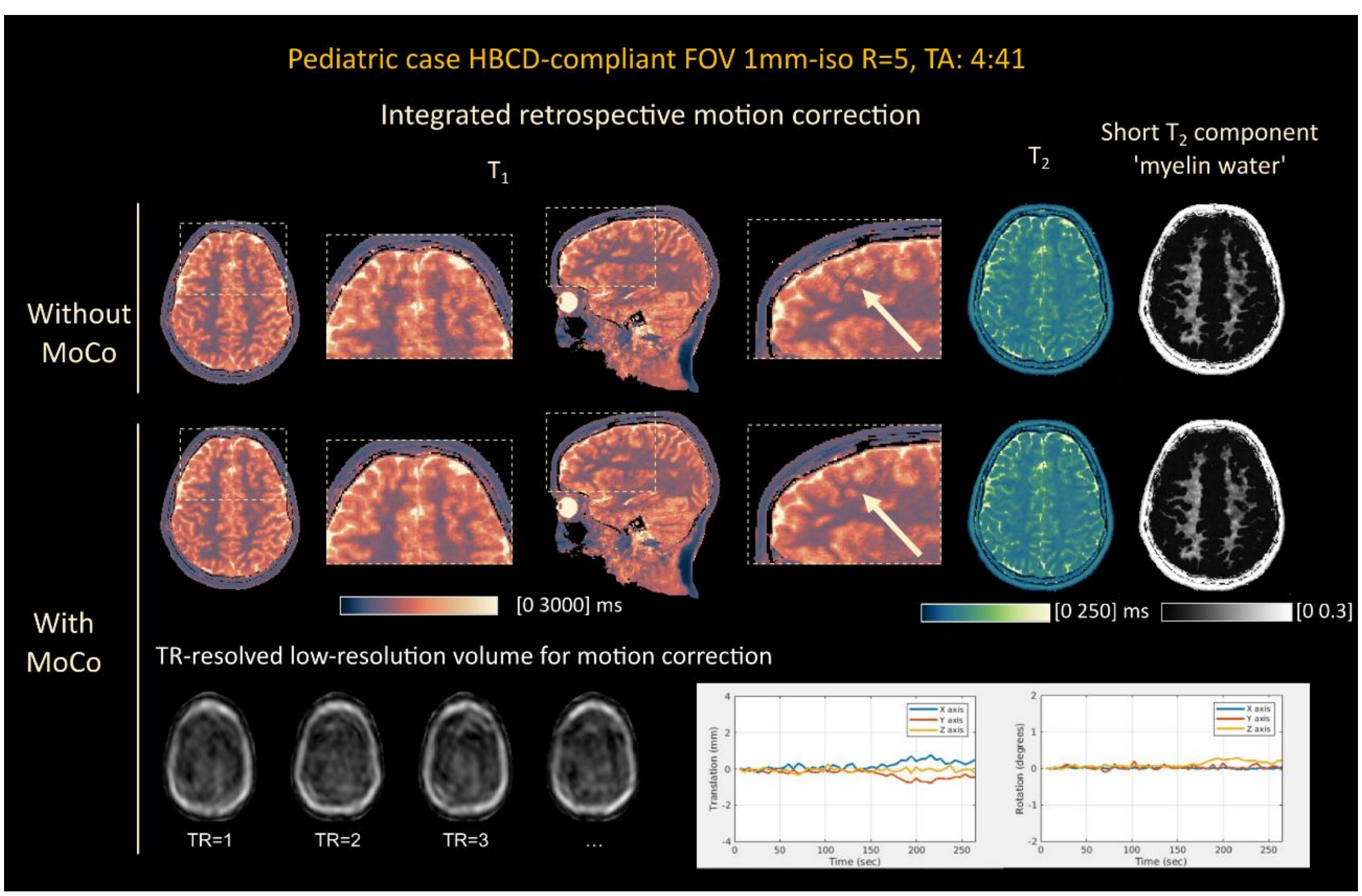

**Figure 7.** Representative pediatric case of a normally developing 12-year-old male subject. Despite the slight motion during the scan as seen in the top row, our integrated retrospective motion correction was able to mitigate the motion artifacts.

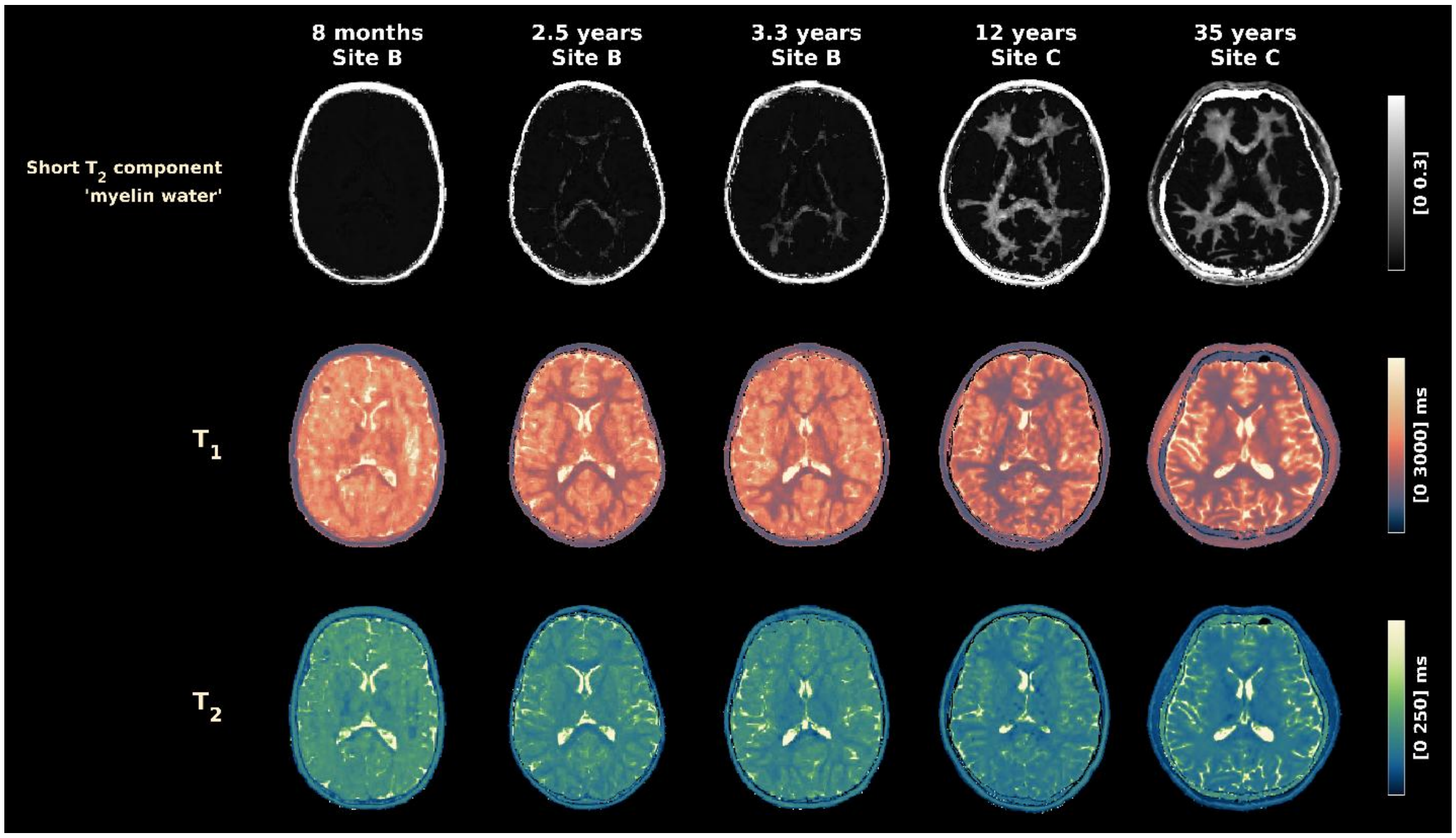

**Figure 8.** MWF maps across pediatric subjects at different developmental stages. $T_1$, $T_2$, and MWF maps from children of different ages demonstrate age-related increases in myelin content across brain regions. The method captures progressive myelination patterns consistent with neurodevelopmental expectations.

## 4. Discussion

This study proposes MWF-QALAS which enables time-efficient, vendor-agnostic myelin-sensitive MR acquisition capable of jointly estimating $T_1$, $T_2$, and MWF maps. Consistent in vivo results were achieved across multiple sites and vendors, demonstrating the technique's potential for clinical and research dissemination. The proposed package provides a valuable tool for quantitative MRI studies of myelin, particularly in multi-site, multi-vendor studies like the HBCD project (Dean et al., 2024).

By incorporating a dual $T_2$-preparation design, retrospective motion correction, a harmonized $B_1^+$ and $B_1^-$ calibration scan, and clinically feasible scan times (4-5 minutes), this method addresses key limitations of current relaxometry techniques, such as long scan times (e.g., IR $T_1$ and SE $T_2$ mapping each take ~20 minutes per slice; whole-brain 3D-GRASE MWF mapping takes 10 minutes at 1.6 $mm^3$ resolution), sensitivity to motion, and lack of cross-vendor compatibility. Through simulations, phantom experiments, and in vivo imaging across Siemens and GE platforms, the sequence demonstrated high quantitative accuracy and reproducibility, particularly in regions with short $T_2$ components that are critical for myelin detection. Furthermore, a pediatric case where visible motion artifacts were largely mitigated demonstrates the feasibility of applying the proposed framework in settings where subject motion is common. While this example is illustrative, the controlled simulation experiments provide the primary quantitative evaluation of motion robustness. The results from pediatric cases suggest that MWF-QALAS may serve as a useful tool for monitoring myelin maturation in longitudinal pediatric studies. These findings position MWF-QALAS as a promising solution for large-scale neurodevelopmental studies such as the HBCD initiative (Dean et al., 2024), which requires harmonized and reproducible imaging protocols across multiple vendors and sites.

Recent technical advances in pediatric myelin imaging have shifted from two-dimensional acquisitions to more time-efficient, volumetric techniques; however, certain limitations persist. Pioneering methods, such as mcDESPOT (Deoni et al., 2012) and two-dimensional MRF approaches (Chen et al., 2019; Li et al., 2019), facilitated quantitative myelin assessment but were confined to 2D encoding with limited slice coverage. In contrast, the 3D ViSTa-MRF approach proposed by Liao et al. (Liao et al., 2024) provides volumetric coverage with increased acquisition efficiency, yet lacks dedicated motion-correction strategies—an essential consideration in pediatric imaging where subject motion is common. For pediatric neurodevelopmental research studies, where sedation is generally not an option, integrated motion correction within a whole-brain, Cartesian, vendor-neutral sequence can be advantageous relative to current MRF implementations. Our technique provides 3D coverage with motion correction at single TR time resolution (5.4 sec) across vendors. These features make the technique feasible for broader clinical adoption and for enabling reliable multi-center investigations.

Although the sequence is open-source and expressed in a vendor-neutral Pulseq (.seq) format, its execution on a scanner requires an installed/enabled vendor-specific Pulseq interpreter as well as local research approvals and site-specific deployment procedures. Interpreters for Siemens and GE systems are available within many research environments (Layton et al., 2017; Nielsen & Noll, 2018), while the GE interpreter is now hosted and distributed by GE HealthCare (https://github.com/GEHC-External/pulseq-ge-interpreter). Philips support has recently been demonstrated and continues to mature (Roos THM et al., MRM 2025; Shaik IA et al., Proc ISMRM 2024). Prototype or vendor-supported implementations have also been reported for additional platforms, including Canon and United Imaging (e.g., via the uMR ADEPT platform). Consequently, the workflow is not immediately "download-and-run": sites must first enable the appropriate interpreter and confirm that the Pulseq events comply with local hardware and safety constraints. Once installed, however, the interpreter is sequence-agnostic and permits execution of the same Pulseq definition across platforms without vendor-specific re-implementation.

While the proposed framework demonstrates strong technical and practical performance, several limitations should be acknowledged. First, the reconstruction is currently performed offline using a MATLAB-based pipeline; no online reconstruction capability has been implemented. This limits feedback at the time of the scan, though future integration with online reconstruction could enable real-time processing directly on the scanner console (Veldmann et al., 2022). Second, while the sequence enables robust detection of myelin-associated short-$T_2$ signals, it is not a direct measurement of myelin. The MWF estimation is based on a three-compartment signal model without exchange (Barta et al., 2015; Chan & Marques, 2020; Deoni et al., 2008) and with fixed relaxation parameters (Chen et al., 2019; Deshmane et al., 2019). Therefore, the accuracy of MWF depends on model assumptions, which can potentially vary across populations or pathologies. Nevertheless, this tradeoff between model simplicity and acquisition efficiency allows for practical whole-brain imaging within minutes while maintaining sensitivity to clinically relevant short-$T_2$ components, as demonstrated by the characterization of normal myelination patterns shown in **Figure 8**. Third, although the method achieves motion-resolved reconstruction at a temporal resolution of ~5.4 s (per TR), finer-scale motion tracking remains limited compared to optical or real-time navigator systems (Zaitsev et al., 2015). Because the pediatric case involved relatively limited motion, additional prospective experiments and retrospective simulations were performed to evaluate performance across a broader range of motion conditions. Within the tested conditions, motion-corrected reconstruction remained effective for translations on the order of a few millimeters and rotations of approximately 1°, including prospective motion reaching approximately 4 mm translation and 1° rotation. These values represent empirical performance within the evaluated range rather than a strict motion tolerance limit, and residual errors increased with motion magnitude, likely because of inaccuracies in inter-TR motion estimation. Our reconstruction/fitting currently assumes small head motion, such that a single B1 map is applicable across the dataset. Given the 4-mm isotropic B1 map used here, a conservative operational bound is that motion moving tissue more than about one B1 voxel relative to the calibration (≈ one voxel translation or the equivalent rotation that displaces tissue by >1 voxel at the brain edge) may introduce bias. We mitigate this by rigidly registering QALAS images to the tAFI volume before fitting. Additionally, reconstruction time remains a bottleneck due to the computational demands of dictionary matching and motion correction, which may benefit from GPU-accelerated, learning-based parameter estimation; for example, SSL-QALAS demonstrated scan-specific fine-tuning using transfer learning within approximately 15 min (Jun et al., 2023). Finally, although we observed agreement with multi-echo GRASE (Piredda et al., 2021), direct histological validation of the estimated MWF was not performed. Accordingly, the term "MWF" in this study should be interpreted as a model-derived estimate rather than a direct measure of myelin. In addition, the comparison between MWF-QALAS and 3D-GRASE was performed in three healthy volunteers, and further evaluation in a larger cohort would help strengthen these findings. Despite these limitations, the combination of short-$T_2$ sensitivity, motion robustness, and vendor-agnostic deployment marks a significant step toward scalable, standardized myelin-sensitive imaging. Future directions include real-time reconstruction implementation, further reduction in scan and reconstruction times, and expanded validation in diverse clinical populations and age groups.

In conclusion, we developed and validated a vendor-agnostic, time-efficient quantitative MRI sequence that enables simultaneous mapping of $T_1$, $T_2$, and MWF within a 4–5 minute scan. By integrating a dual $T_2$-preparation scheme, retrospective motion correction, and harmonized $B_1^+$ calibration, the proposed method addresses key limitations in existing myelin imaging techniques, including long acquisition times, poor motion tolerance, and vendor-specific constraints. Validation through simulation, phantom, and in vivo studies demonstrated the feasibility of integrating MWF-QALAS into large-scale neuroimaging studies and pave the way for standardized, motion-robust, and reproducible myelin-sensitive imaging across diverse populations and imaging platforms.

## 5. Data and Code Availability

The pulse sequence and reconstruction codes can be found here: https://github.com/unaydorken/mwf-qalas.

## 6. Author Contributions

The authors confirm contribution to the paper as follows: study conception and design: SF, YR, BG, BB; supervision of research: BG, BB; data collection: UDG, SF, YJ, CA, YC, XY, QL, SY, OA, CJ, BG, BB; analysis of data: UDG, SF, YJ, BG, BB; technical support: YJ, ADK, GFP, TH, EM, QL, SY, MZ, JFN; interpretation of results: YJ, KSC, YR, MZ, JFN, OA, CJ, PEG, BG, BB; draft manuscript preparation: UDG, SF, BG, BB. All authors reviewed the results, edited the manuscript, and approved its final version.

## 7. Funding

This work was supported by research grants NIH R01 EB028797, U01 EB025162, P41 EB030006, U01 EB026996, R03 EB031175, R01 EB032378, UG3 EB034875, U01 DA055353, R01 EB034757, R01 NS133228, R01 NS121657, JSPS Overseas Research Fellowship, and NVidia Corporation for computing support.

## 8. Declaration of Competing Interests

Eugene Milshteyn is currently employed at GE HealthCare. Antoine Delattre-Klauser, Gian Franco Piredda, and Tom Hilbert are currently employed at Siemens Healthineers.

## 9. Supplementary Material

Supplementary material for this article is available with the online version.

## References

Afacan, O., Erem, B., Roby, D. P., Roth, N., Roth, A., Prabhu, S. P., & Warfield, S. K. (2016). Evaluation of motion and its effect on brain magnetic resonance image quality in children. *Pediatric Radiology*, *46*(12), 1728–1735.

A G Teixeira, R. P., Neji, R., Wood, T. C., Baburamani, A. A., Malik, S. J., & Hajnal, J. V. (2020). Controlled saturation magnetization transfer for reproducible multivendor variable flip angle T1 and T2 mapping. *Magnetic Resonance in Medicine*, *84*(1), 221–236.

Andre, J. B., Bresnahan, B. W., Mossa-Basha, M., Hoff, M. N., Smith, C. P., Anzai, Y., & Cohen, W. A. (2015). Toward quantifying the prevalence, severity, and cost associated with patient motion during clinical MR examinations. *Journal of the*

*American College of Radiology: JACR*, *12*(7), 689–695.
Barta, R., Kalantari, S., Laule, C., Vavasour, I. M., MacKay, A. L., & Michal, C. A. (2015). Modeling T(1) and T(2) relaxation in bovine white matter. *Journal of Magnetic Resonance (San Diego, Calif.: 1997)*, *259*, 56–67.
Boudreau, M., Karakuzu, A., Cohen-Adad, J., Bozkurt, E., Carr, M., Castellaro, M., Concha, L., Doneva, M., Dual, S. A., Ensworth, A., Foias, A., Fortier, V., Gabr, R. E., Gilbert, G., Glide-Hurst, C. K., Grech-Sollars, M., Hu, S., Jalnefjord, O., Jovicich, J., … ISMRM Reproducible Research Study Group and the ISMRM Quantitative MR Study Group. (2024). Repeat it without me: Crowdsourcing the T1 mapping common ground via the ISMRM reproducibility challenge. *Magnetic Resonance in Medicine*, *92*(3), 1115–1127.
Boudreau, M., Tardif, C. L., Stikov, N., Sled, J. G., Lee, W., & Pike, G. B. (2017). B1 mapping for bias-correction in quantitative T1 imaging of the brain at 3T using standard pulse sequences. *Journal of Magnetic Resonance Imaging*, *46*(6), 1673–1682.
Cao, X., Liao, C., Iyer, S. S., Wang, Z., Zhou, Z., Dai, E., Liberman, G., Dong, Z., Gong, T., He, H., Zhong, J., Bilgic, B., & Setsompop, K. (2022). Optimized multi-axis spiral projection MR fingerprinting with subspace reconstruction for rapid whole-brain high-isotropic-resolution quantitative imaging. *Magnetic Resonance in Medicine*, *88*(1), 133–150.
Casey, B. J., Cannonier, T., Conley, M. I., Cohen, A. O., Barch, D. M., Heitzeg, M. M., Soules, M. E., Teslovich, T., Dellarco, D. V., Garavan, H., Orr, C. A., Wager, T. D., Banich, M. T., Speer, N. K., Sutherland, M. T., Riedel, M. C., Dick, A. S., Bjork, J. M., Thomas, K. M., … ABCD Imaging Acquisition Workgroup. (2018). The Adolescent Brain Cognitive Development (ABCD) study: Imaging acquisition across 21 sites. *Developmental Cognitive Neuroscience*, *32*, 43–54.
Carter, F., Anwander, A., Johnson, M., Goucha, T., Adamson, H., Friederici, A. D., Lutti, A., Gauthier, C. J., Weiskopf, N., Bazin, P.-L., & Steele, C. J. (2025). Assessing quantitative MRI techniques using multimodal comparisons. *PLOS ONE, 20*(7), e0327828.
Chan, K.-S., & Marques, J. P. (2020). Multi-compartment relaxometry and diffusion informed myelin water imaging - Promises and challenges of new gradient echo myelin water imaging methods. *NeuroImage*, *221*(117159), 117159.
Chen, Y., Chen, M.-H., Baluyot, K. R., Potts, T. M., Jimenez, J., Lin, W., & UNC/UMN Baby Connectome Project Consortium. (2019). MR fingerprinting enables quantitative measures of brain tissue relaxation times and myelin water fraction in the first five years of life. *NeuroImage*, *186*, 782–793.

Cho, J., Gagoski, B., Kim, T. H., Wang, F., Manhard, M. K., Dean, D., 3rd, Kecskemeti, S., Caprihan, A., Lo, W.-C., Splitthoff, D. N., Liu, W., Polak, D., Cauley, S., Setsompop, K., Grant, P. E., & Bilgic, B. (2024). Time-efficient, high-resolution 3T whole-brain relaxometry using 3D-QALAS with wave-CAIPI readouts. *Magnetic Resonance in Medicine*, *91*(2), 630–639.

Christodoulou, A. G., Shaw, J. L., Nguyen, C., Yang, Q., Xie, Y., Wang, N., & Li, D. (2018). Magnetic resonance multitasking for motion-resolved quantitative cardiovascular imaging. *Nature Biomedical Engineering*, *2*(4), 215–226.

Dean, D. C., 3rd, Tisdall, M. D., Wisnowski, J. L., Feczko, E., Gagoski, B., Alexander, A. L., Edden, R. A. E., Gao, W., Hendrickson, T. J., Howell, B. R., Huang, H., Humphreys, K. L., Riggins, T., Sylvester, C. M., Weldon, K. B., Yacoub, E., Ahtam, B., Beck, N., Banerjee, S., … HBCD MRI Working Group. (2024). Quantifying brain development in the HEALthy Brain and Child Development (HBCD) Study: The magnetic resonance imaging and spectroscopy protocol. *Developmental Cognitive Neuroscience*, *70*(101452), 101452.

Deoni, S. C. L., Dean, D. C., 3rd, O'Muircheartaigh, J., Dirks, H., & Jerskey, B. A. (2012). Investigating white matter development in infancy and early childhood using myelin water faction and relaxation time mapping. *NeuroImage*, *63*(3), 1038–1053.

Deoni, S. C. L., Rutt, B. K., Arun, T., Pierpaoli, C., & Jones, D. K. (2008). Gleaning multicomponent T1 and T2 information from steady-state imaging data: 2D Relaxometry With Steady-State Imaging. *Magnetic Resonance in Medicine*, *60*(6), 1372–1387.

Deshmane, A., McGivney, D. F., Ma, D., Jiang, Y., Badve, C., Gulani, V., Seiberlich, N., & Griswold, M. A. (2019). Partial volume mapping using magnetic resonance fingerprinting. *NMR in Biomedicine*, *32*(5), e4082.

Frahm, J., Haase, A., & Matthaei, D. (1986). Rapid NMR imaging of dynamic processes using the FLASH technique. *Magnetic Resonance in Medicine*, *3*(2), 321–327.

Fujita, S., Gagoski, B., Hwang, K.-P., Hagiwara, A., Warntjes, M., Fukunaga, I., Uchida, W., Saito, Y., Sekine, T., Tachibana, R., Muroi, T., Akatsu, T., Kasahara, A., Sato, R., Ueyama, T., Andica, C., Kamagata, K., Amemiya, S., Takao, H., … Aoki, S. (2024). Cross-vendor multiparametric mapping of the human brain using 3D-QALAS: A multicenter and multivendor study. *Magnetic Resonance in Medicine*, *91*(5), 1863–1875.

Fujita, S., Hagiwara, A., Hori, M., Warntjes, M., Kamagata, K., Fukunaga, I., Andica, C., Maekawa, T., Irie, R., Takemura, M. Y., Kumamaru, K. K., Wada, A., Suzuki, M., Ozaki, Y., Abe, O., & Aoki, S. (2019). Three-dimensional high-resolution simultaneous quantitative mapping of the whole brain with 3D-QALAS: An accuracy and repeatability study. *Magnetic Resonance Imaging*, *63*, 235–243.

Gallichan, D., Marques, J. P., & Gruetter, R. (2016). Retrospective correction of involuntary microscopic head movement using highly accelerated fat image navigators (3D FatNavs) at 7T. *Magnetic Resonance in Medicine*, 75(3), 1030–1039.

Gómez, P. A., Cencini, M., Golbabaee, M., Schulte, R. F., Pirkl, C., Horvath, I., Fallo, G., Peretti, L., Tosetti, M., Menze, B. H., & Buonincontri, G. (2020). Rapid three-dimensional multiparametric MRI with quantitative transient-state imaging. *Scientific Reports*, *10*(1), 13769.

Griesler, T., Stebani, J., Kaplan, S., Angelov, I., Albert, P., Blaimer, M., Wech, T., Wang, X., Chen, Q., Zaitsev, M., Zhu, Z., Liu, Q., Martin, P., Nielsen, J.-F., Hamilton, J. I., Nordbeck, P., Seiberlich, N., & Gram, M. (2026). OpenMRF: A modular, vendor-neutral open-source framework for reproducible magnetic resonance fingerprinting using Pulseq. arXiv preprint arXiv:2604.22713.

Haase, A., Frahm, J., Matthaei, D., Hanicke, W., & Merboldt, K.-D. (1986). FLASH imaging. Rapid NMR imaging using low flip-angle pulses. *Journal of Magnetic Resonance*, *67*(2), 258–266.

Henschel, L., Conjeti, S., Estrada, S., Diers, K., Fischl, B., & Reuter, M. (2020). FastSurfer - A fast and accurate deep learning based neuroimaging pipeline. *NeuroImage*, *219*, 117012.

Howell, B. R., Styner, M. A., Gao, W., Yap, P.-T., Wang, L., Baluyot, K., Yacoub, E., Chen, G., Potts, T., Salzwedel, A., Li, G., Gilmore, J. H., Piven, J., Smith, J. K., Shen, D., Ugurbil, K., Zhu, H., Lin, W., & Elison, J. T. (2019). The UNC/UMN Baby Connectome Project (BCP): An overview of the study design and protocol development. *NeuroImage*, *185*, 891–905.

Jenista, E. R., Rehwald, W. G., Chen, E.-L., Kim, H. W., Klem, I., Parker, M. A., & Kim, R. J. (2013). Motion and flow insensitive adiabatic T2 -preparation module for cardiac MR imaging at 3 Tesla. *Magnetic Resonance in Medicine: Official Journal of the Society of Magnetic Resonance in Medicine / Society of Magnetic Resonance in Medicine*, *70*(5), 1360–1368.

Jiang, Y., Ma, D., Keenan, K. E., Stupic, K. F., Gulani, V., & Griswold, M. A. (2017). Repeatability of magnetic resonance fingerprinting T1 and T2 estimates assessed using the ISMRM/NIST MRI system phantom. *Magnetic Resonance in Medicine*, *78*(4), 1452–1457.

Jun, Y., Cho, J., Wang, X., Gee, M., Grant, P. E., Bilgic, B., & Gagoski, B. (2023). SSL-QALAS: Self-supervised learning for rapid multiparameter estimation in quantitative MRI using 3D-QALAS. *Magnetic Resonance in Medicine, 90*(5), 2019–2032.

Karakuzu, A., Biswas, L., Cohen-Adad, J., & Stikov, N. (2022). Vendor-neutral sequences and fully transparent workflows improve inter-vendor reproducibility of quantitative MRI. *Magnetic Resonance in Medicine*, *88*(3), 1212–1228.

Kvernby, S., Warntjes, M., Carlhäll, C.-J., Engvall, J., & Ebbers, T. (2014). 3D-Quantification using an interleaved Look-Locker acquisition sequence with T2-prep pulse (3D-QALAS). *Journal of Cardiovascular Magnetic Resonance: Official Journal of the Society for Cardiovascular Magnetic Resonance*, *16*(S1), O82.

Labadie, C., Lee, J.-H., Rooney, W. D., Jarchow, S., Aubert-Frécon, M., Springer, C. S., Jr, & Möller, H. E. (2014). Myelin water mapping by spatially regularized longitudinal relaxographic imaging at high magnetic fields. *Magnetic Resonance in Medicine*, *71*(1), 375–387.

Laule, C., Vavasour, I. M., Kolind, S. H., Li, D. K. B., Traboulsee, T. L., Moore, G. R. W., & MacKay, A. L. (2007). Magnetic resonance imaging of myelin. *Neurotherapeutics: The Journal of the American Society for Experimental NeuroTherapeutics*, *4*(3), 460–484.

Layton, K. J., Kroboth, S., Jia, F., Littin, S., Yu, H., Leupold, J., Nielsen, J.-F., Stöcker, T., & Zaitsev, M. (2017). Pulseq: A rapid and hardware-independent pulse sequence prototyping framework. *Magnetic Resonance in Medicine*, *77*(4), 1544–1552.

Lebel, C., & Deoni, S. (2018). The development of brain white matter microstructure. *NeuroImage*, *182*, 207–218.

Lee, J., Hyun, J.-W., Lee, J., Choi, E.-J., Shin, H.-G., Min, K., Nam, Y., Kim, H. J., & Oh, S.-H. (2021). So you want to image myelin using MRI: An overview and practical guide for myelin water imaging. *Journal of Magnetic Resonance Imaging*, *53*(2), 360–373.

Liao, C., Cao, X., Iyer, S. S., Schauman, S., Zhou, Z., Yan, X., Chen, Q., Li, Z., Wang, N., Gong, T., Wu, Z., He, H., Zhong, J., Yang, Y., Kerr, A., Grill-Spector, K., & Setsompop, K. (2024). High-resolution myelin-water fraction and quantitative relaxation mapping using 3D ViSTa-MR fingerprinting. *Magnetic Resonance in Medicine*, *91*(6), 2278–2293.

Li, Q., Cao, X., Ye, H., Liao, C., He, H., & Zhong, J. (2019). Ultrashort echo time magnetic resonance fingerprinting (UTE-MRF) for simultaneous quantification of long and ultrashort T2 tissues. *Magnetic Resonance in Medicine*, *82*(4), 1359–1372.

National Institute on Drug Abuse. (2021, December 2). *The HEALthy brain and child*

*development study*. National Institute on Drug Abuse. https://nida.nih.gov/research/nida-research-programs-activities/healthy-brain-child-development-study

Nave, K.-A. (2010). Myelination and support of axonal integrity by glia. *Nature*, *468*(7321), 244–252.

Nielsen, J.-F., & Noll, D. C. (2018). TOPPE: A framework for rapid prototyping of MR pulse sequences. *Magnetic Resonance in Medicine*, *79*(6), 3128–3134.

Paus, T., Keshavan, M., & Giedd, J. N. (2008). Why do many psychiatric disorders emerge during adolescence? *Nature Reviews. Neuroscience*, *9*(12), 947–957.

Piredda, G. F., Hilbert, T., Canales-Rodríguez, E. J., Pizzolato, M., von Deuster, C., Meuli, R., Pfeuffer, J., Daducci, A., Thiran, J.-P., & Kober, T. (2021). Fast and high-resolution myelin water imaging: Accelerating multi-echo GRASE with CAIPIRINHA. *Magnetic Resonance in Medicine*, *85*(1), 209–222.

Pruessmann, K. P., Weiger, M., Scheidegger, M. B., & Boesiger, P. (1999). SENSE: sensitivity encoding for fast MRI. *Magnetic Resonance in Medicine*, *42*(5), 952–962.

Saltarelli, G., Di Cerbo, G., Innocenzi, A., De Felici, C., Splendiani, A., & Di Cesare, E. (2025). Quantitative MRI in neuroimaging: A review of techniques, biomarkers, and emerging clinical applications. *Brain Sciences, 15*(10), 1088.

Seiberlich, N., Gulani, V., Campbell-Washburn, A., Sourbron, S., Doneva, M. I., Calamante, F., & Hu, H. H. (2020). *Quantitative Magnetic Resonance Imaging*. Academic Press.

Shrout, P. E., & Fleiss, J. L. (1979). Intraclass correlations: uses in assessing rater reliability. *Psychological Bulletin*, *86*(2), 420–428.

Somerville, L. H., Bookheimer, S. Y., Buckner, R. L., Burgess, G. C., Curtiss, S. W., Dapretto, M., Elam, J. S., Gaffrey, M. S., Harms, M. P., Hodge, C., Kandala, S., Kastman, E. K., Nichols, T. E., Schlaggar, B. L., Smith, S. M., Thomas, K. M., Yacoub, E., Van Essen, D. C., & Barch, D. M. (2018). The Lifespan Human Connectome Project in Development: A large-scale study of brain connectivity development in 5-21 year olds. *NeuroImage*, *183*, 456–468.

Tamir, J. I., Uecker, M., Chen, W., Lai, P., Alley, M. T., Vasanawala, S. S., & Lustig, M. (2017). T2 shuffling: Sharp, multicontrast, volumetric fast spin-echo imaging. *Magnetic Resonance in Medicine*, 77(1), 180–195. https://doi.org/10.1002/mrm.26102

Tofts, P. (2005). *Quantitative MRI of the Brain: Measuring Changes Caused by Disease*. John Wiley & Sons.

Uecker, M., Ong, F., Tamir, J. I., Bahri, D., Virtue, P., Cheng, J. Y., Zhang, T., & Lustig,

M. (2015). Berkeley Advanced Reconstruction Toolbox. *Proceedings of the International Society for Magnetic Resonance in Medicine*, 23, 2486.

Veldmann, M., Ehses, P., Chow, K., Nielsen, J.-F., Zaitsev, M., & Stöcker, T. (2022). Open-source MR imaging and reconstruction workflow. *Magnetic Resonance in Medicine*, *88*(6), 2395–2407.

Wang, F., Dong, Z., Reese, T. G., Rosen, B., Wald, L. L., & Setsompop, K. (2022). 3D Echo Planar Time-resolved Imaging (3D-EPTI) for ultrafast multi-parametric quantitative MRI. *NeuroImage*, *250*(118963), 118963.

Weiner, M. W., Veitch, D. P., Aisen, P. S., Beckett, L. A., Cairns, N. J., Green, R. C., Harvey, D., Jack, C. R., Jr, Jagust, W., Morris, J. C., Petersen, R. C., Salazar, J., Saykin, A. J., Shaw, L. M., Toga, A. W., Trojanowski, J. Q., & Alzheimer's Disease Neuroimaging Initiative. (2017). The Alzheimer's Disease Neuroimaging Initiative 3: Continued innovation for clinical trial improvement. *Alzheimer's & Dementia: The Journal of the Alzheimer's Association*, *13*(5), 561–571.

Weiskopf, N., Suckling, J., Williams, G., Correia, M. M., Inkster, B., Tait, R., Ooi, C., Bullmore, E. T., & Lutti, A. (2013). Quantitative multi-parameter mapping of R1, PD*, MT, and R2* at 3T: A multi-center validation. *Frontiers in Neuroscience, 7*, 95.

Wicaksono, K. P., Fushimi, Y., Nakajima, S., Sakata, A., Okuchi, S., Hinoda, T., Oshima, S., Otani, S., Tagawa, H., Urushibata, Y., & Nakamoto, Y. (2023). Accuracy, repeatability, and reproducibility of T1 and T2 relaxation times measurement by 3D magnetic resonance fingerprinting with different dictionary resolutions. *European Radiology*, *33*(4), 2895–2904.

Yarnykh, V. L. (2007). Actual flip-angle imaging in the pulsed steady state: a method for rapid three-dimensional mapping of the transmitted radiofrequency field. *Magnetic Resonance in Medicine*, *57*(1), 192–200.

Zaitsev, M., Maclaren, J., & Herbst, M. (2015). Motion artifacts in MRI: A complex problem with many partial solutions. *Journal of Magnetic Resonance Imaging*, *42*(4), 887–901.

**Supplementary Table 1:** Acquisition parameters of inversion recovery $T_1$ and single-echo spin echo $T_2$ mapping

| | Inversion-recovery fast-spin-echo | single-echo fast-spin-echo |
|---|---|---|
| Field of view (mm) | 192 × 192 | 192 × 192 |
| Matrix Size | 192 × 192 | 192 × 192 |
| Slice Thickness (mm) | 3 | 3 |
| Band width (Hz/pixel) | 338 | 338 |
| Repetition time (s) | 8.1 | 1.5 |
| Echo time (ms) | 7.6 | 10.0 |
| Inversion time (ms) | [35, 100, 150, 250, 500, 1000, 2000, 3000, 4000] | [10, 30, 50, 70, 90, 120, 200, 300, 400] |
| Turbo factor | 18 | 80 |
| Acceleration | 2 | 3 |
| Scan time | 2 min 36 sec per TI | 2 min 3 sec per TE |

Note–. ISMRM/NIST: International Society for Magnetic Resonance in Medicine and National Institute of Standards and Technology

**Supplementary Table 2:** Brain structures evaluated in vivo

| Structure | Tissue type |
|---|---|
| Left cerebral white matter | White matter |
| Left cerebral cortex | Gray matter |
| Left lateral ventricle | Cerebrospinal fluid |
| Left inferior lateral ventricle | Cerebrospinal fluid |
| Left cerebellum white matter | White matter |
| Left cerebellum cortex | Gray matter |
| Left thalamus | Gray matter |
| Left caudate | Gray matter |
| Left putamen | Gray matter |
| Left pallidum | Gray matter |
| 3rd ventricle | Cerebrospinal fluid |
| 4th ventricle | Cerebrospinal fluid |
| Brain-stem | Gray matter |
| Left hippocampus | Gray matter |
| Left amygdala | Gray matter |
| Left accumbens area | Gray matter |
| Left ventral DC | Gray matter |
| Right cerebral white matter | White matter |
| Right cerebral cortex | Gray matter |
| Right lateral ventricle | Cerebrospinal fluid |
| Right inferior lateral ventricle | Cerebrospinal fluid |
| Right cerebellum white matter | White matter |
| Right cerebellum cortex | Gray matter |
| Right thalamus | Gray matter |
| Right caudate | Gray matter |
| Right putamen | Gray matter |
| Right pallidum | Gray matter |
| Right hippocampus | Gray matter |
| Right amygdala | Gray matter |
| Right accumbens area | Gray matter |
| Right ventral DC | Gray matter |

Note–. DC, diencephalon.

**Supplementary Table 3.** Absolute error across the full range of $T_1$ and $T_2$ values. Values are reported as median (25th, 75th percentile) due to non-normal distributions. Absolute error is defined as the absolute difference between estimated and ground truth values.

| Metric | Noise level | QALAS absolute error (ms) | MWF-QALAS absolute error (ms) |
|---|---|---|---|
| $T_1$ | 1% | 40.8 (12.6, 113.2) | 42.1 (13.6, 115.3) |
| | 3% | 111.8 (34.2, 286.4) | 114.6 (36.7, 290.5) |
| | 5% | 174.4 (55.5, 421.5) | 178.4 (58.6, 430.8) |
| $T_2$ | 1% | 5.1 (2.1, 11.8) | 4.7 (1.6, 12.3) |
| | 3% | 13.5 (5.6, 28.2) | 12.6 (4.2, 31.1) |
| | 5% | 20.6 (8.8, 42.1) | 19.7 (6.6, 46.1) |

**Supplementary Table 4.** Absolute errors for $T_1$ and $T_2$ estimation using QALAS and MWF-QALAS within a restricted range ($T_1$ < 300 ms and $T_2$ < 30 ms) across different noise levels (1%, 3%, and 5%). Values are reported as median (25th, 75th percentile) due to non-normal distributions. Absolute error is defined as the absolute difference between estimated and reference values.

| Metric | Noise level | QALAS absolute error (ms) | MWF-QALAS absolute error (ms) |
|---|---|---|---|
| $T_1$ | 1% | 6.5 (2.2, 27.5) | 10.3 (4.2, 56.4) |
| | 3% | 15.0 (5.0, 44.4) | 18.1 (7.0, 74.4) |
| | 5% | 20.9 (7.5, 60.0) | 25.2 (8.9, 89.0) |
| $T_2$ | 1% | 6.1 (2.5, 12.3) | 0.8 (0.3, 1.4) |
| | 3% | 12.1 (5.9, 19.4) | 1.9 (0.9, 3.4) |
| | 5% | 16.3 (8.2, 24.6) | 3.0 (1.4, 5.4) |

**Supplementary Table 5.** Pairwise cross-vendor comparison of quantitative $T_1$, $T_2$, and myelin water fraction (MWF) measurements across Siemens Prisma, Siemens Skyra, and GE Premier scanners. Pearson correlation coefficients together with Bland–Altman mean differences (bias) and 95% limits of agreement (LoA) are reported using pooled structure-wise measurements from three healthy adult subjects.

| Parameter | Scanner comparison | Pearson r | Mean difference | Lower 95% LoA | Upper 95% LoA |
|---|---|---|---|---|---|
| **$T_1$ (ms)** | Siemens Prisma − Siemens Skyra | 0.921 | −74.21 ms | −759.37 ms | 610.94 ms |
| **$T_1$ (ms)** | Siemens Prisma − GE Premier | 0.965 | −216.75 ms | −800.20 ms | 366.70 ms |
| **$T_1$ (ms)** | Siemens Skyra − GE Premier | 0.976 | −142.54 ms | −654.87 ms | 369.79 ms |
| **$T_2$ (ms)** | Siemens Prisma − Siemens Skyra | 0.988 | −0.071 ms | −29.22 ms | 29.08 ms |
| **$T_2$ (ms)** | Siemens Prisma − GE Premier | 0.991 | −0.731 ms | −28.33 ms | 26.86 ms |
| **$T_2$ (ms)** | Siemens Skyra − GE Premier | 0.986 | −0.661 ms | −35.96 ms | 34.64 ms |
| **MWF** | Siemens Prisma − Siemens Skyra | 0.706 | 0.0041 | −0.0490 | 0.0572 |
| **MWF** | Siemens Prisma − GE Premier | 0.888 | 0.0104 | −0.0265 | 0.0472 |
| **MWF** | Siemens Skyra − GE Premier | 0.902 | 0.0062 | −0.0219 | 0.0344 |

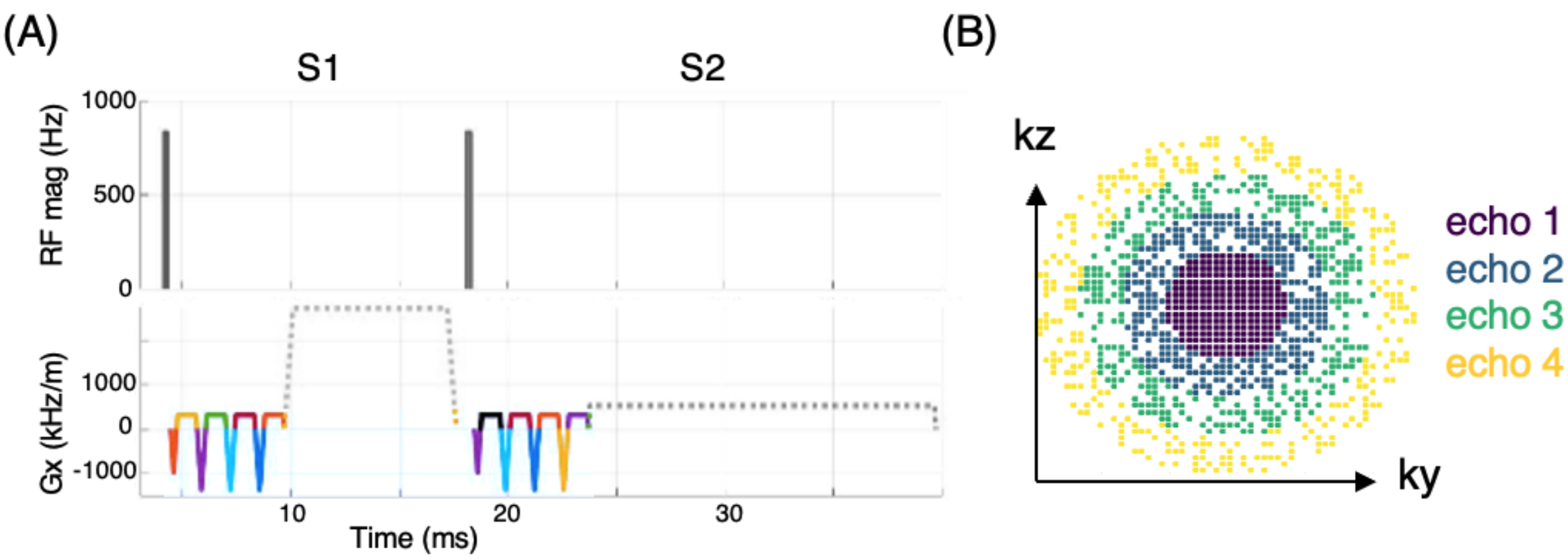

**Supplementary Figure 1.** AFI is extended by incorporating turbo readout (echo train length = 4). (A) Pulse sequence diagram of turbo AFI. (B) Sampling pattern of turbo AFI. Turbo factor of 4 was used and sampling was performed in a center-out fashion. AFI, actual flip angle imaging.

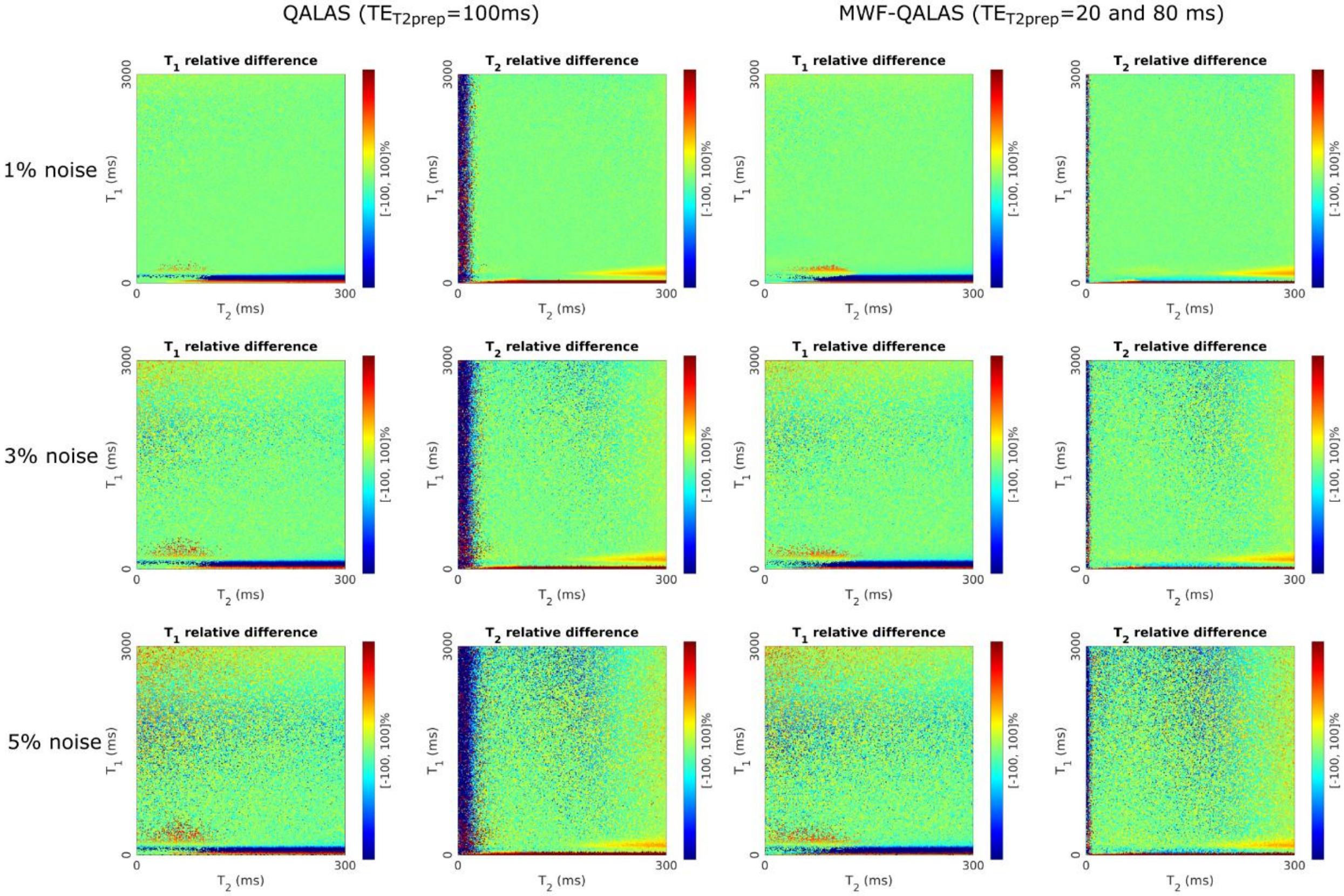

**Supplementary Figure 2. Effect of noise level on $T_1$ and $T_2$ estimation accuracy.** Relative-difference maps are shown for standard QALAS with a single $T_2$-preparation time of 100 ms and MWF-QALAS with $T_2$-preparation times of 20 and 80 ms. Rows correspond to Gaussian noise levels of 1%, 3%, and 5%, while columns show $T_1$ and $T_2$ relative differences for each method. Relative difference was calculated as 100×(reference−estimated)/reference, with a display range of −100% to 100%. MWF-QALAS demonstrates reduced $T_2$ estimation error in the short-$T_2$ range across all evaluated noise levels.

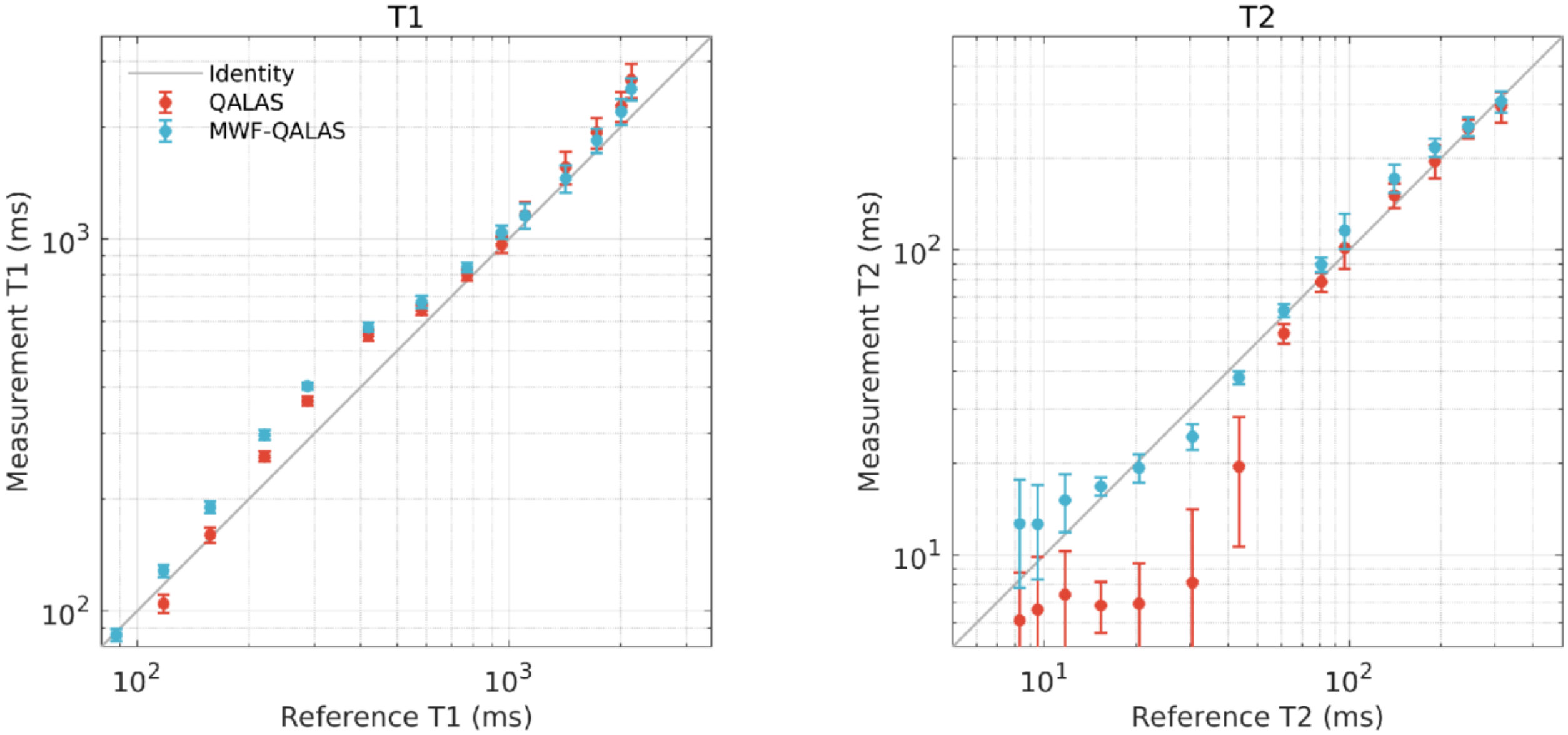

**Supplementary Figure 3.** $T_1$ and $T_2$ Mapping Validation in NIST/ISMRM Phantom. Quantitative $T_1$ and $T_2$ maps obtained with reference methods (inversion recovery for $T_1$ and spin echo for $T_2$), QALAS, and MWF-QALAS. Scatter plots of measured versus reference $T_1$ and $T_2$ values for the QALAS (red) and MWF- QALAS (blue) methods. Values displayed in log-log plot.

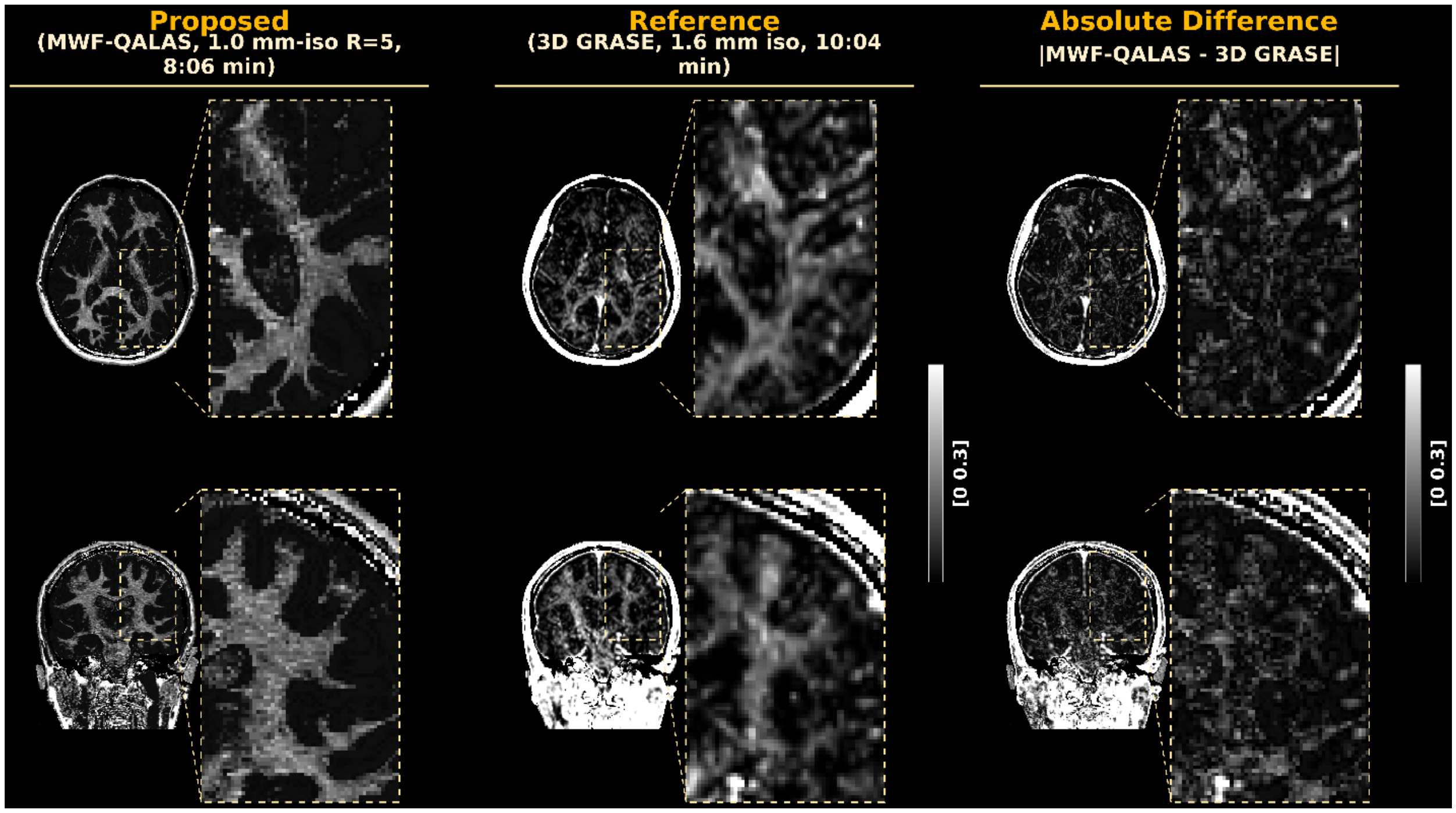

**Supplementary Figure 4.** Comparison of myelin water fraction (MWF) maps between the proposed MWF-QALAS method and the reference 3D-GRASE technique. Axial (top) and coronal (bottom) MWF maps are shown for a healthy adult subject. Left panels: MWF-QALAS maps acquired at 1.0 mm isotropic resolution. Middle panels: Reference 3D-GRASE maps acquired at 1.6 mm isotropic resolution in 10 minutes and 30 seconds. Right panels: Voxel-wise absolute difference between methods. The MWF values are displayed on a shared grayscale color bar ranging from 0 to 0.3.

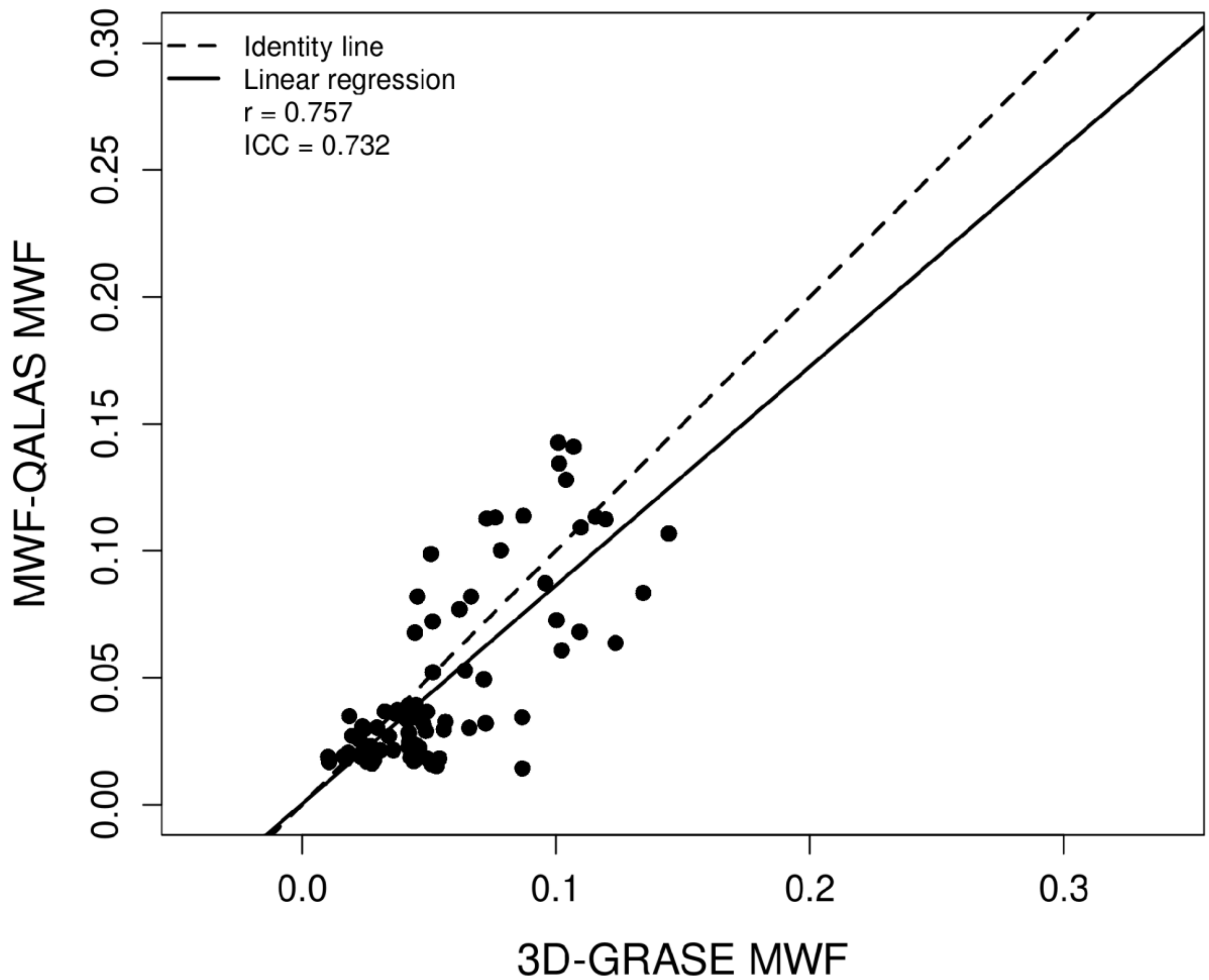

**Supplementary Figure 5.** Structure-wise comparison of MWF measurements obtained using MWF-QALAS and 3D-GRASE. Each point represents one anatomical structure from one of three healthy adult subjects. The solid line indicates the linear regression fit, and the dashed line indicates the identity line. The comparison yielded Pearson's correlation coefficient of r=0.757 and an intraclass correlation coefficient of 0.732.

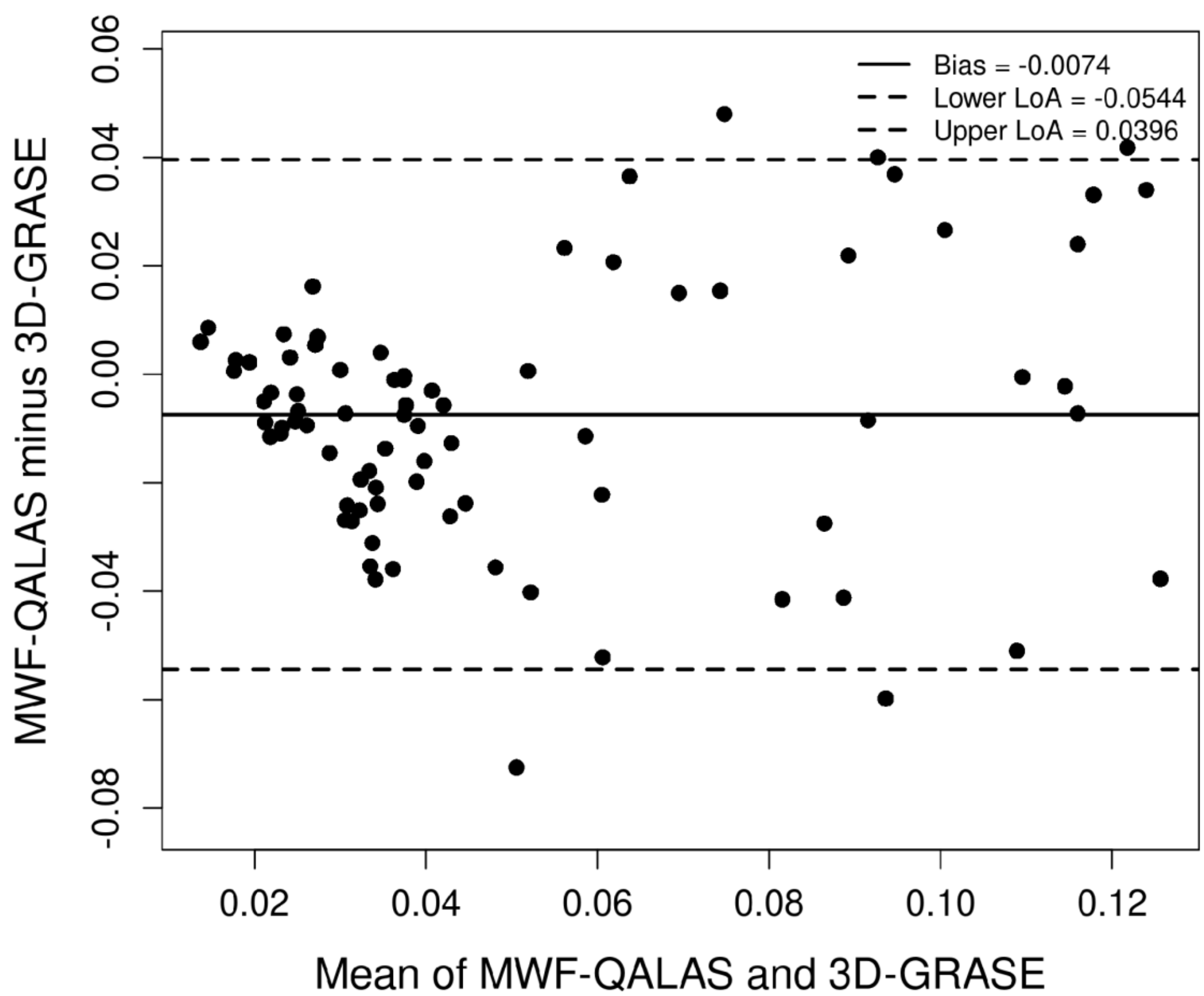

**Supplementary Figure 6.** Bland–Altman analysis of MWF measurements obtained using MWF-QALAS and 3D-GRASE. Each point represents one anatomical structure from one of three healthy adult subjects. The solid line indicates the mean difference between MWF-QALAS and 3D-GRASE of −0.0074, and the dashed lines indicate the limits of agreement from −0.0544 to 0.0396. Negative differences indicate higher MWF measurements with 3D-GRASE than with MWF-QALAS.

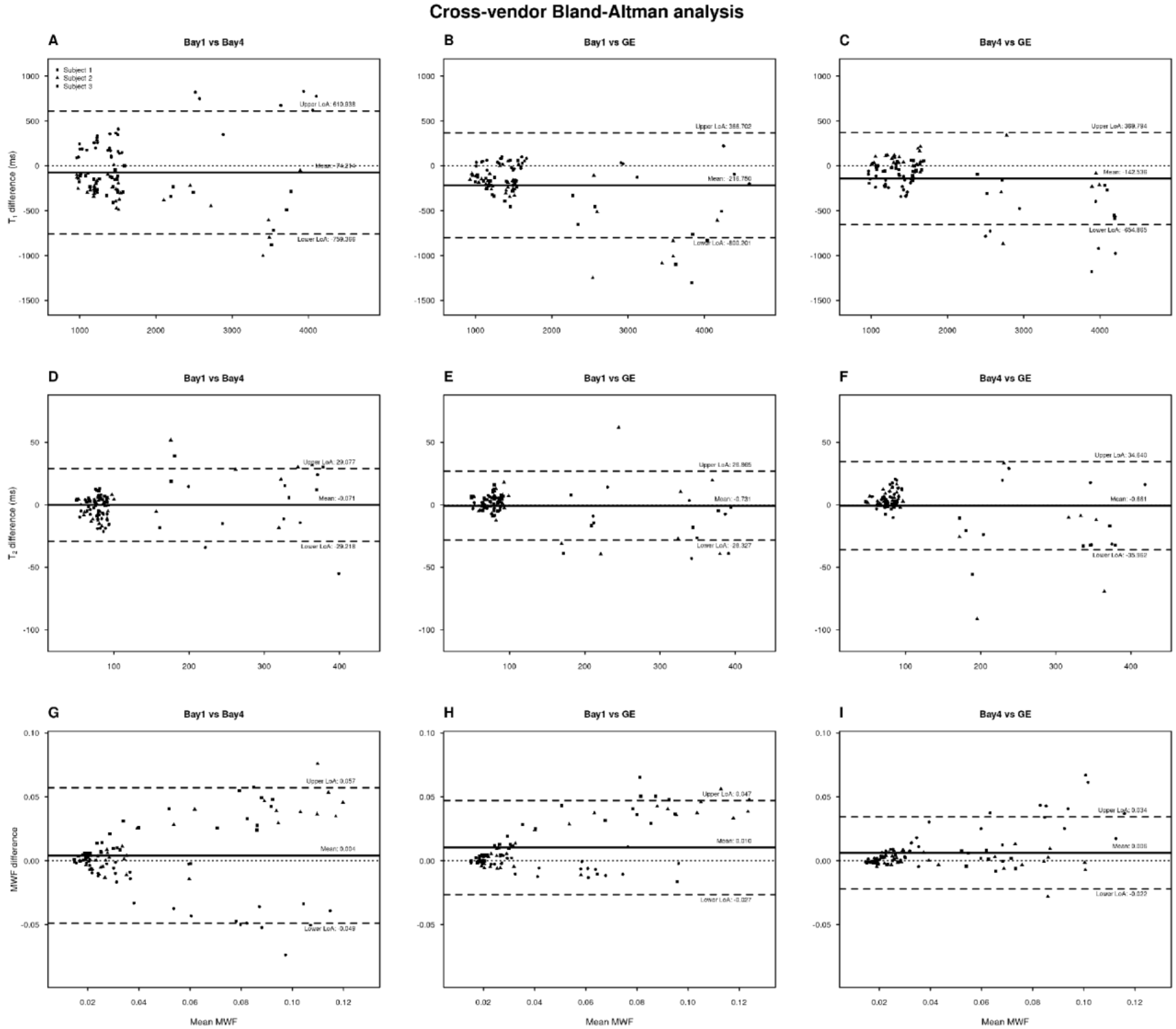

**Supplementary Figure 7.** Bland–Altman analysis of cross-vendor agreement for quantitative $T_1$, $T_2$, and myelin water fraction (MWF) measurements across three healthy adult subjects. Rows correspond to $T_1$, $T_2$, and MWF, and columns show pairwise comparisons between Siemens Prisma and Siemens Skyra, Siemens Prisma and GE Premier, and Siemens Skyra and GE Premier. The solid line indicates the mean difference (bias), and the dashed lines indicate the 95% limits of agreement (mean ± 1.96 SD). Different marker shapes represent the three subjects.

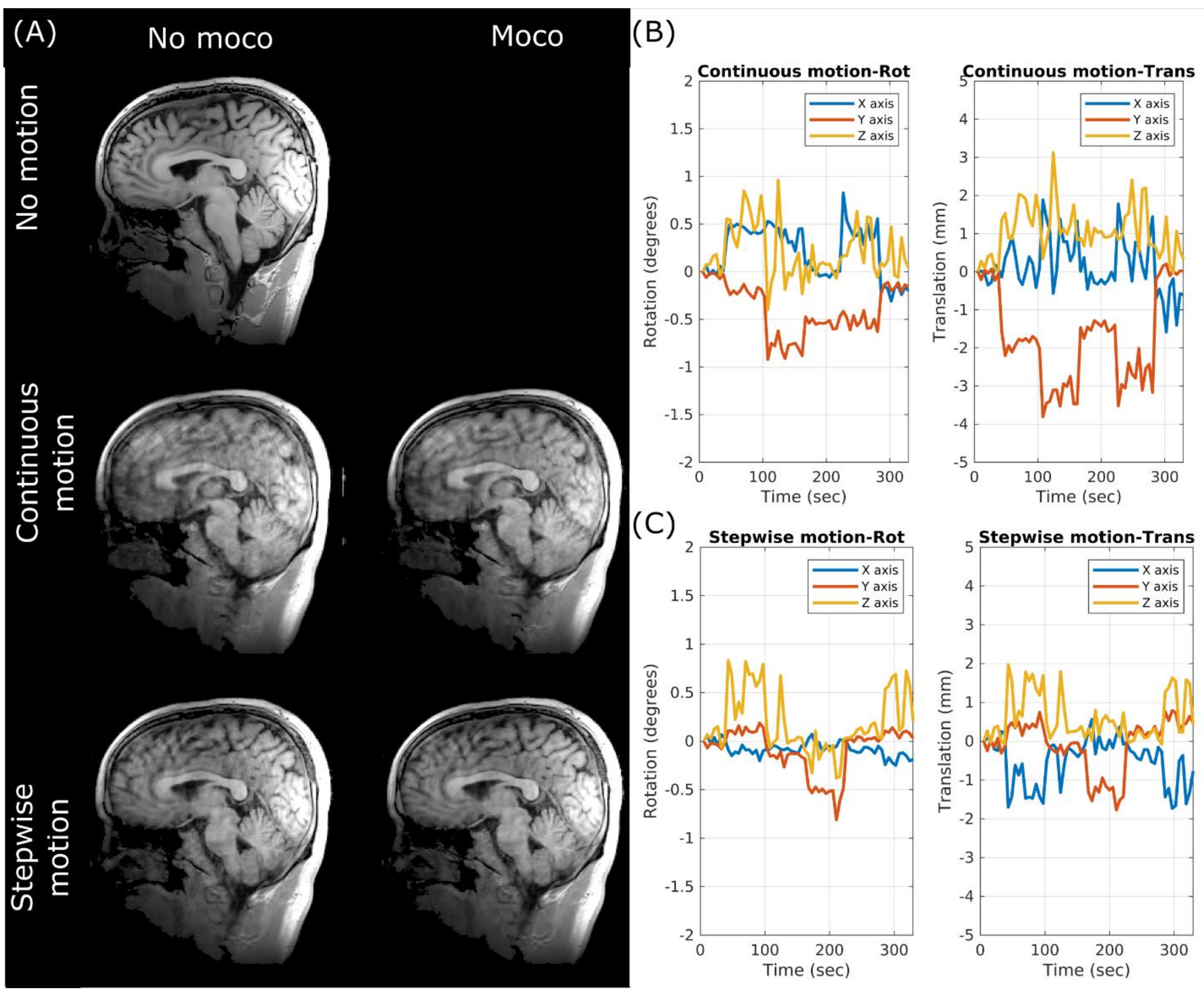

**Supplementary Figure 8.** Prospective motion correction validation. (A) Representative reconstructions of the fourth MWF-QALAS contrast from the no-motion acquisition (Experiment 1), continuous-motion acquisition (Experiment 2), and stepwise-motion acquisition (Experiment 3), shown without and with motion correction where applicable. Only the fourth contrast is displayed to simplify visual comparison. (B) Estimated rotational and translational motion trajectories for Experiment 2. (C) Estimated rotational and translational motion trajectories for Experiment 3.

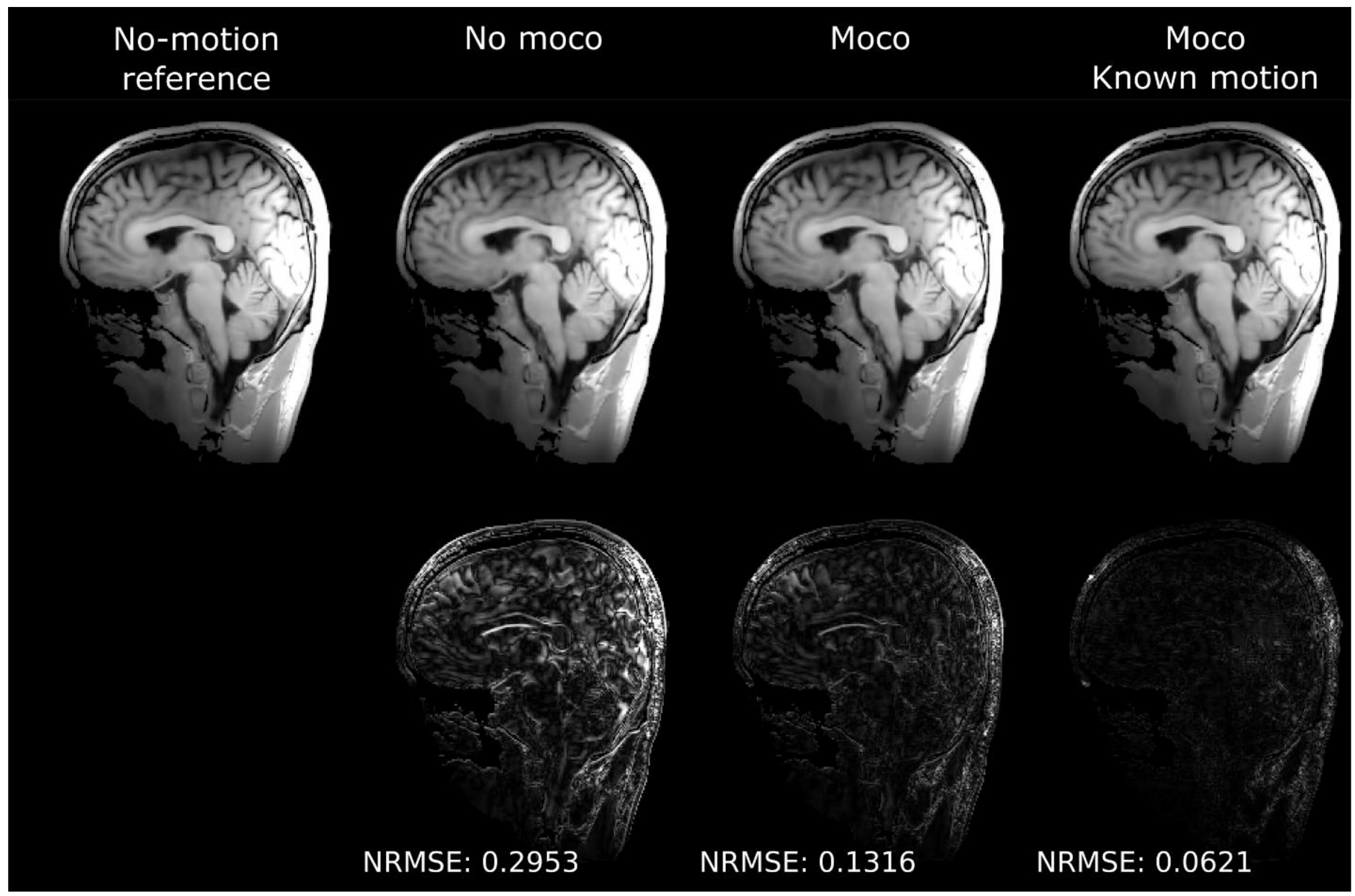

**Supplementary Figure 9.** Retrospective motion correction simulation using the prospectively estimated motion trajectory (Experiment 4). The motion trajectory estimated from the prospective continuous-motion acquisition was applied to the motion-free dataset to generate motion-corrupted data. The top row shows the motion-free reference and reconstructions obtained without motion correction, with motion correction using the estimated trajectory, and with motion correction using the known applied trajectory. A representative reconstruction of the fourth MWF-QALAS contrast is displayed. The bottom row shows the corresponding absolute difference maps relative to the motion-free reference. The NRMSE was reduced from 29.53% without motion correction to 13.16% using the estimated trajectory and 6.21% using the known applied trajectory.

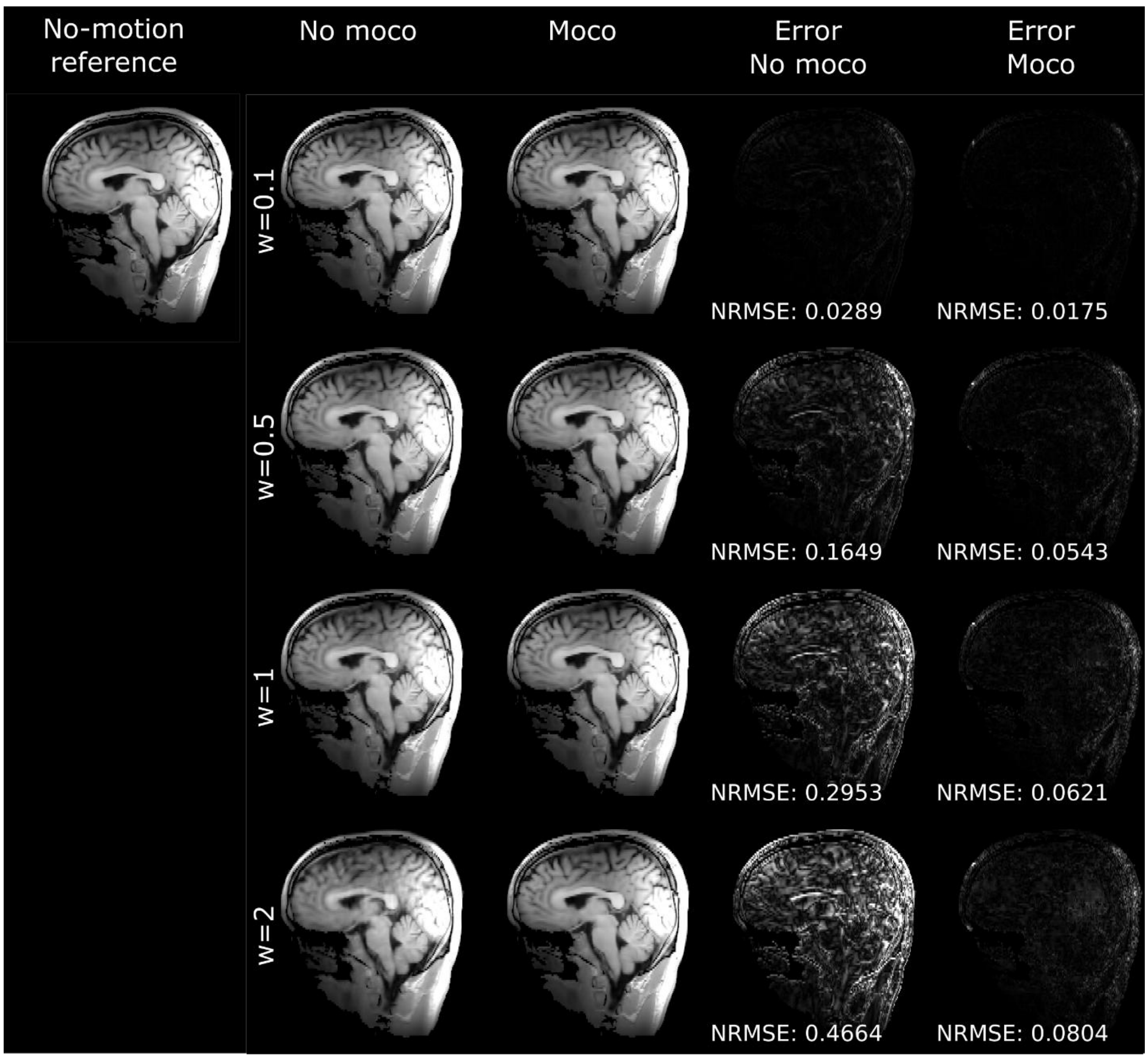

**Supplementary Figure 10.** Effect of motion magnitude on retrospective motion correction (Experiment 5). The motion trajectory used in Experiment 4 was scaled by factors of w= 0.1, 0.5, 1, and 2 and applied to the motion-free data. Representative reconstructions of the fourth MWF-QALAS contrast are shown without motion correction and with motion correction, together with the corresponding absolute difference maps relative to the motion-free reference. Motion correction reduced the NRMSE from 2.89% to 1.75%, from 16.49% to 5.43%, from 29.53% to 6.21%, and from 46.64% to 8.04% for w=0.1, 0.5, 1, and 2, respectively.

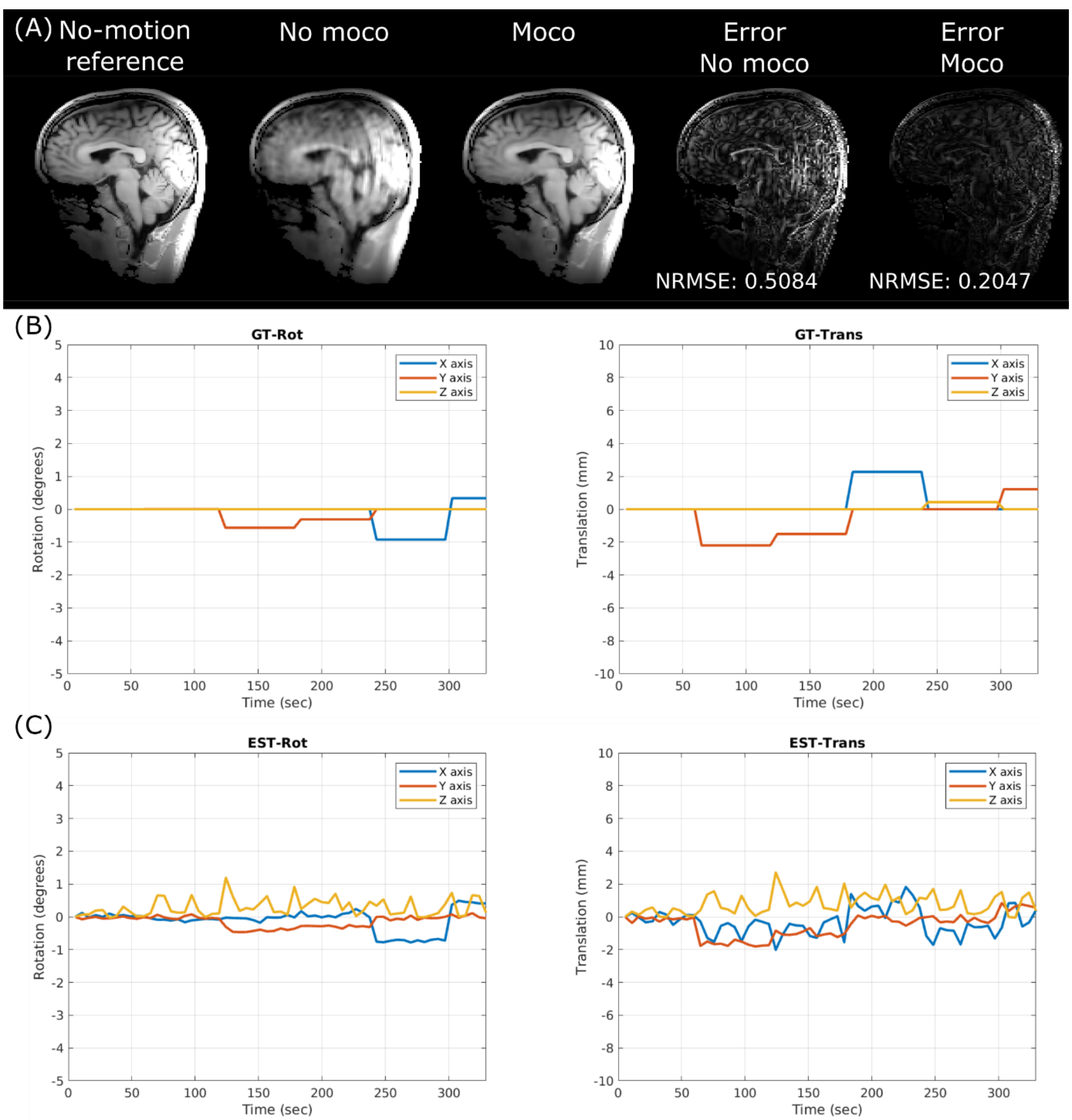

**Supplementary Figure 11.** Retrospective motion correction with a randomly varying stepwise motion trajectory (Experiment 6). (A) Representative reconstruction of the fourth MWF-QALAS contrast from the motion-free reference, the simulated motion-corrupted data without motion correction, and the reconstruction after motion correction using the estimated trajectory. The corresponding absolute difference maps relative to the motion-free reference are shown. Motion correction reduced the NRMSE from 50.84% to 20.47%. (B) Known applied rotational and translational motion trajectories used to generate the motion-corrupted data. The motion was updated approximately every 1 min, with maximum rotations of 1° and translations of 3 mm. (C) Rotational and translational motion trajectories estimated from the simulated motion-corrupted data.

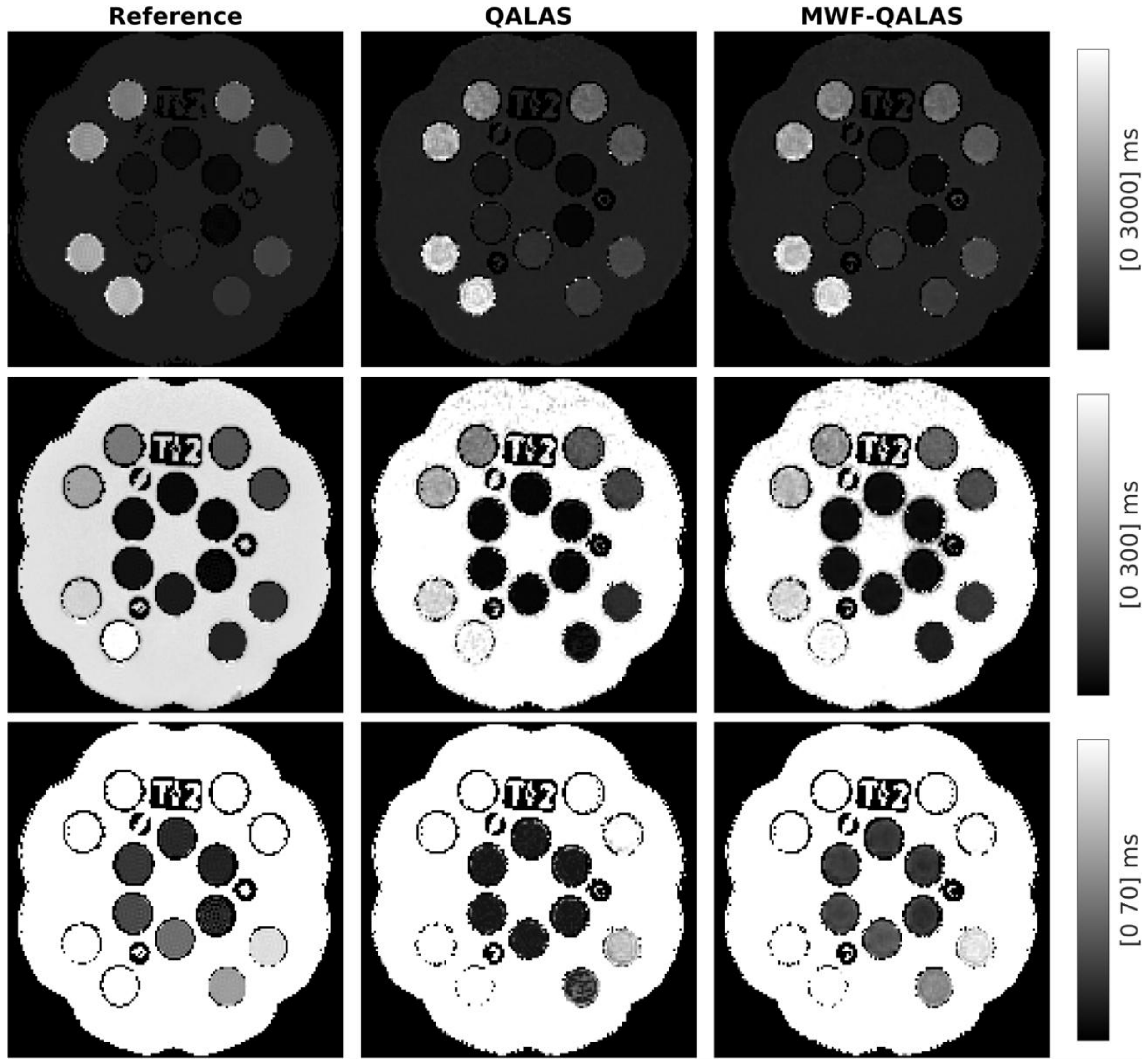

**Supplementary Figure 12.** Grayscale version of Figure 4. Quantitative $T_1$ and $T_2$ maps obtained with reference methods (inversion recovery for $T_1$ and spin echo for $T_2$), QALAS, and MWF-QALAS. Lower-range $T_2$ scale displays shorter $T_2$ values relevant for myelin (bottom row).

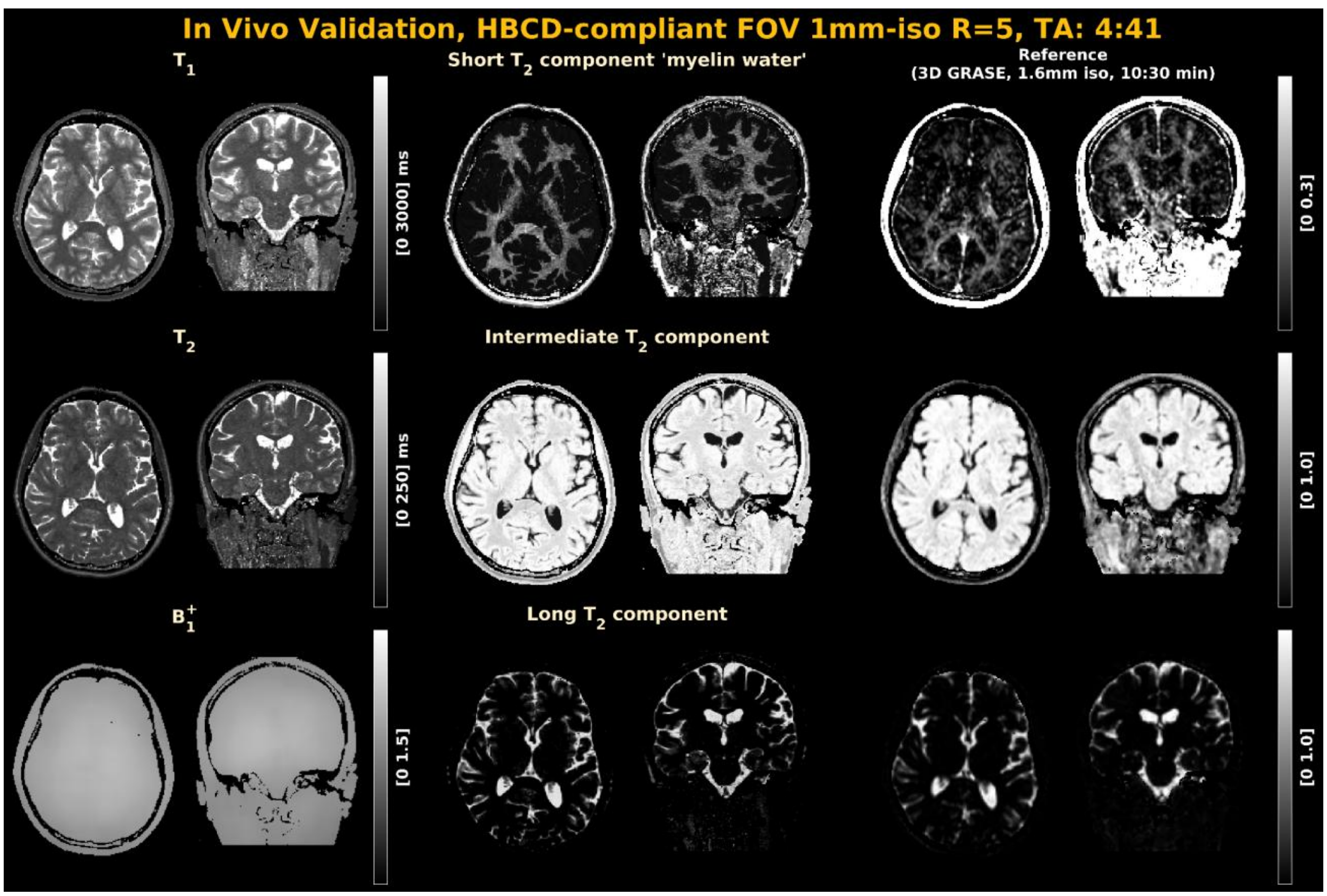

**Supplementary Figure 13.** Grayscale version of Figure 5 showing representative in vivo $T_1$, $T_2$, $B_1^+$, and tissue-compartment maps obtained with MWF-QALAS, together with the corresponding 3D-GRASE reference maps.

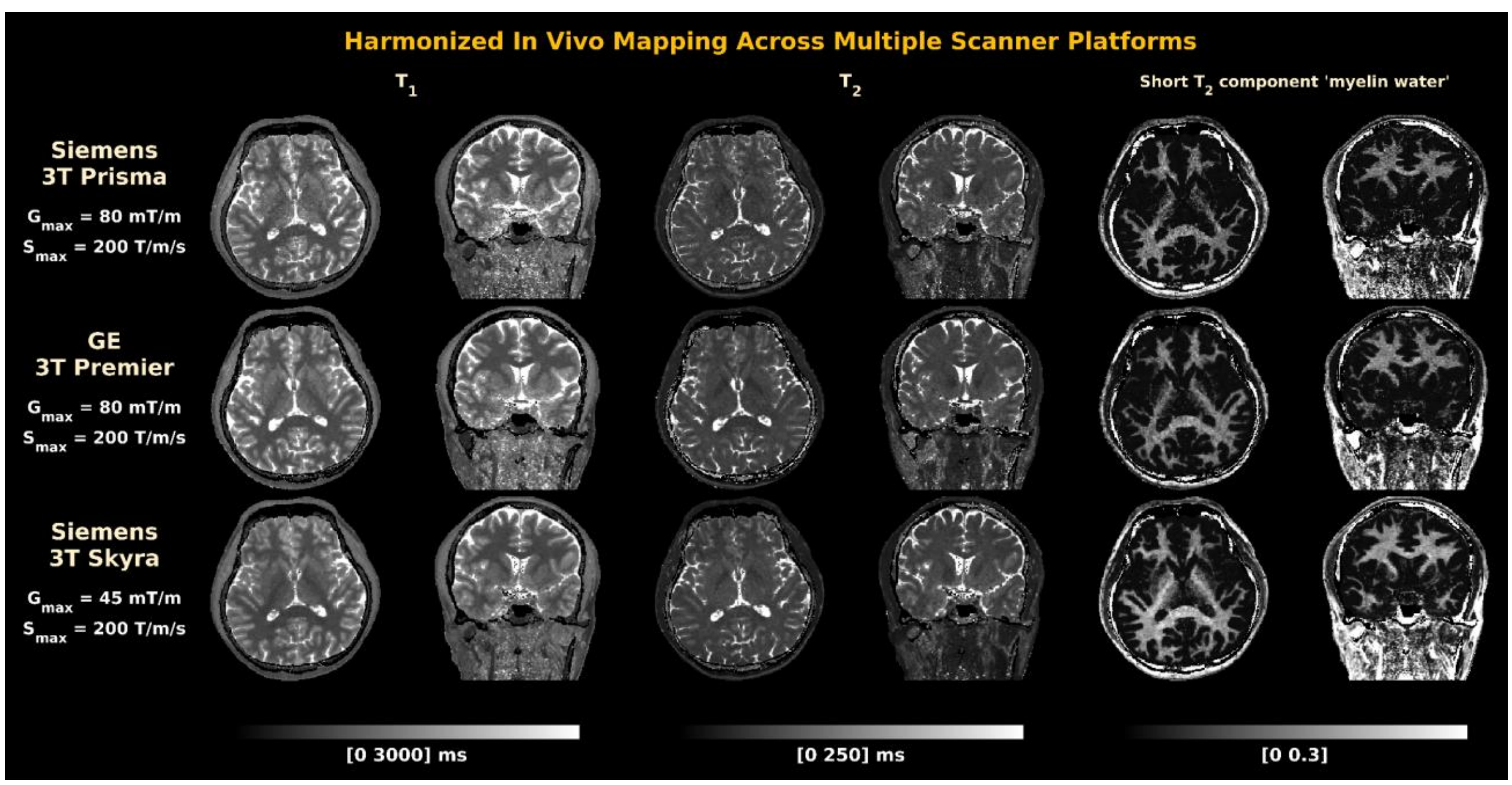

**Supplementary Figure 14.** Grayscale version of Figure 6 showing representative $T_1$, $T_2$, and myelin water fraction maps acquired across Siemens Prisma, GE Premier, and Siemens Skyra scanners.

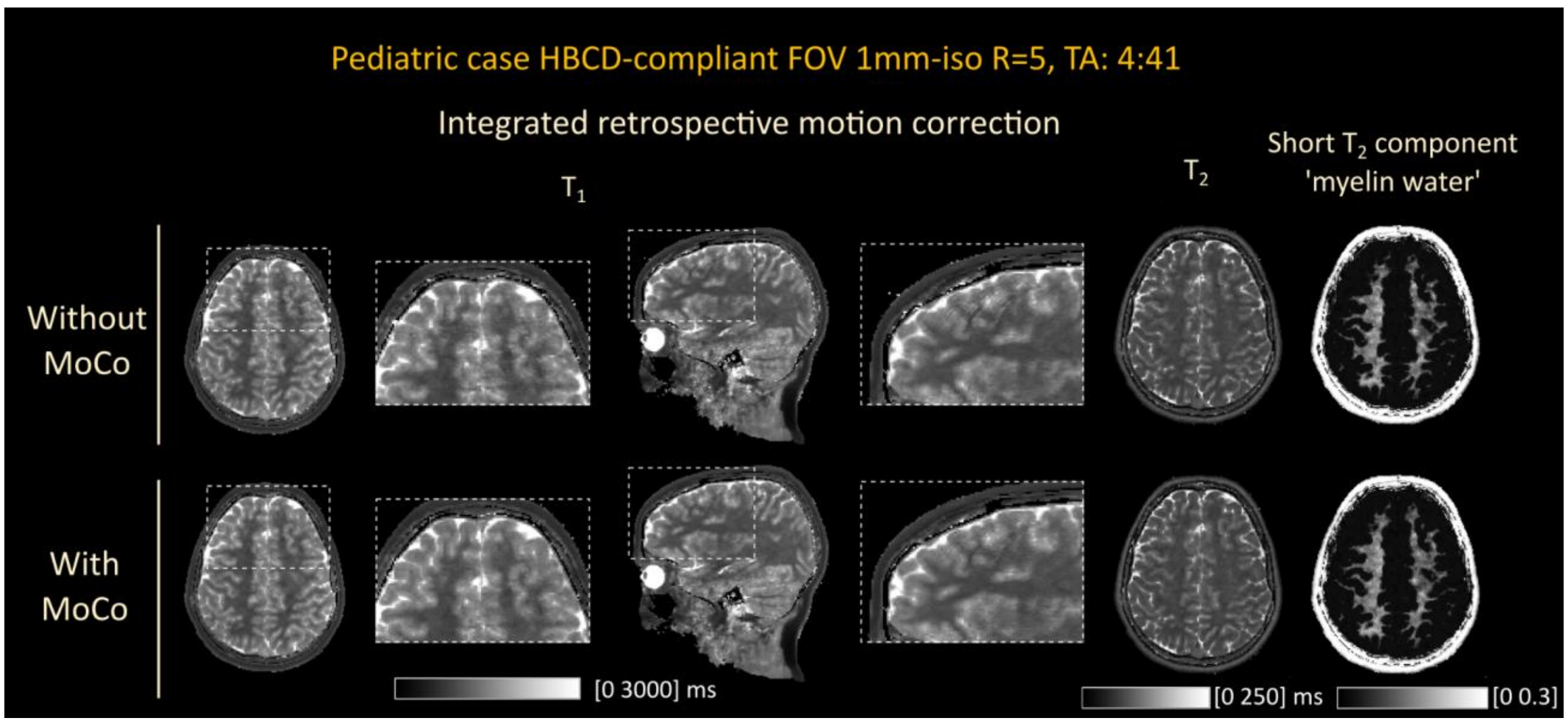

**Supplementary Figure 15.** Grayscale version of Figure 7 showing representative $T_1$, $T_2$, and myelin water fraction maps before and after motion correction in the pediatric in vivo dataset, including the corresponding zoomed views.

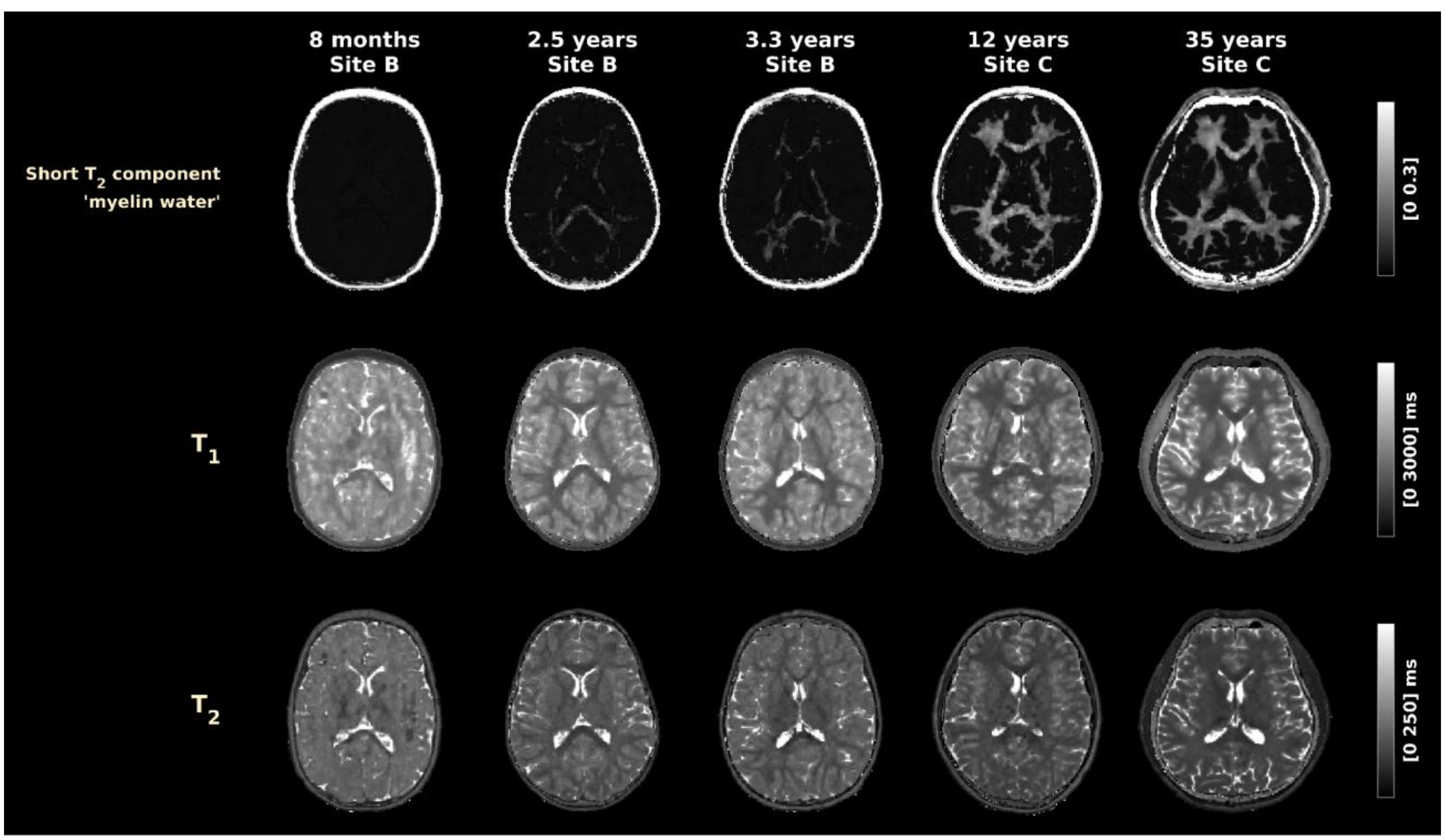

**Supplementary Figure 16.** Grayscale version of Figure 8 showing representative myelin water fraction, $T_1$, and $T_2$ maps from pediatric subjects of different ages and an adult volunteer.